\documentclass[twocolumn]{aastex63}

\def\cof {$^{12}\mathrm{CO}~(J=1\rightarrow0)$}
\def\cos {$^{13}\mathrm{CO}~(J=1\rightarrow0)$}
 \def\cotss{$^{12}\mathrm{CO}~(J=3\rightarrow2)$}

\def\cot {$\mathrm{C}^{18}\mathrm{O}~(J=1\rightarrow0)$}
\def\msun{$M_{\odot}$}
\def\hh{$\mathrm{H}_2$} 
\def\ag{$A_{G}$}
\def\av{$A_{V}$}
\def\cofs {$^{12}\mathrm{CO}$}
\def\coss {$^{13}\mathrm{CO}$}
\def\cots{$\mathrm{C}^{18}\mathrm{O}$}
\def\deg  {\ifmmode {^\circ}\else {$^\circ$}\fi}

\def\kms     {km~s$^{-1}$}
\newcommand{\HII}{\mbox{H\,\textsc{ii}}}%

\usepackage{booktabs}
\usepackage{chngcntr}

\date{\today}
\submitjournal{ApJ}

\shorttitle{Local molecular clouds}
\shortauthors{Yan et al.} 

\begin{document}

\title{Distances and statistics of local molecular clouds in the first Galactic quadrant}

\correspondingauthor{Ji Yang}
\email{jiyang@pmo.ac.cn,qzyan@pmo.ac.cn}

\author[0000-0003-4586-7751]{Qing-Zeng Yan}
\affil{Purple Mountain Observatory and Key Laboratory of Radio Astronomy,\\
 Chinese Academy of Sciences, Nanjing 210034, People's Republic of China}

\author[0000-0001-7768-7320]{Ji Yang}
\affil{Purple Mountain Observatory and Key Laboratory of Radio Astronomy,\\
 Chinese Academy of Sciences, Nanjing 210034, People's Republic of China}

 \author[0000-0002-0197-470X]{Yang Su }
 \affil{Purple Mountain Observatory and Key Laboratory of Radio Astronomy,\\
  Chinese Academy of Sciences, Nanjing 210034, People's Republic of China}

 \author[0000-0002-3904-1622]{Yan Sun}
 \affil{Purple Mountain Observatory and Key Laboratory of Radio Astronomy,\\
  Chinese Academy of Sciences, Nanjing 210034, People's Republic of China}
 
  \author[0000-0001-8923-7757]{Chen Wang}
 \affil{Purple Mountain Observatory and Key Laboratory of Radio Astronomy,\\
  Chinese Academy of Sciences, Nanjing 210034, People's Republic of China}





\begin{abstract}
We present an analysis of local molecular clouds ($-6 <V_{\rm LSR}< 30$ \kms, i.e., $<1.5$ kpc) in the first Galactic quadrant ($25.8\deg <l<49.7\deg$ and $|b|<5\deg$), a pilot region of the Milky Way Imaging Scroll Painting (MWISP) CO survey. Using the SCIMES algorithm to divide large molecular clouds into moderate-sized ones, we determined distances to 28 molecular clouds with the background-eliminated extinction-parallax (BEEP) method using the $Gaia$ DR2 parallax measurements aided by \ag\ and \av, and the distance ranges from 250 pc to about 1.5 kpc. These incomplete distance samples indicate a linear relationship between the distance and the radial velocity ($V_{\rm LSR}$) with a scatter of 0.16 kpc, and kinematic distances may be systematically larger for local molecular clouds. In order to investigate fundamental properties of molecular clouds, such as the total sample number, the linewidth, the brightness temperature, the physical area, and the mass, we decompose the spectral cube using the DBSCAN algorithm. Post selection criteria are imposed on DBSCAN clusters to remove the noise contamination, and we found that the separation of molecular cloud individuals is reliable based on a definition of independent consecutive structures in $l$-$b$-$V$ space.  The completeness of the local molecular cloud flux collected by the MWISP CO survey is about 80\%. The physical area, $A$, shows a power-law distribution, d$N$/d$A \propto A^{-2.20\pm0.18}$, while the molecular cloud mass also follows a power-law distribution but slightly flatter, d$N$/d$M \propto M^{-1.96\pm0.11}$. 
\end{abstract}



\section{Introduction} \label{sec:intro}


Molecular clouds, the molecular phase of the interstellar medium (ISM), are fundamental components of the Milky Way \citep{1937ApJ....86..483S,1963Natur.200..829W, 1970ApJ...161L..81C,2015ARA&A..53..583H}. The sizes and masses of molecular clouds range  over several orders of magnitudes \citep{2018ARA&A..56...41M}, from small diffuse molecular clouds \citep[e.g., those at high Galactic latitudes with sizes and masses of 0.1 pc and 0.1 \msun, respectively,][]{1996ApJS..106..447M} to giant molecular clouds \citep[40 pc and $10^5$ \msun,][]{1979IAUS...84...35S,1980gmcg.work...41S}. The radial velocity dispersion (the Gaussian standard deviation of the spectral profile) of molecular clouds is about 1-10 \kms\ \citep{1970ApJ...161L..43W,1981MNRAS.194..809L, 2019A&A...628A..78R}. Although molecular clouds are involved in a number of physical processes occurring in the Milky Way, such as star formation \citep[e.g.,][]{2007ARA&A..45..565M,2012ARA&A..50..531K} and the large-scale Galactic structure development \citep[e.g.,][]{1986ApJ...305..892D,2010ApJ...723..492R}, many key mechanism of molecular clouds are still unclear, for instance, their formation and evolution \citep[see][for a review]{2014prpl.conf....3D}.





Large-scale CO surveys, with high spatial dynamic ranges, are essential to understanding molecular clouds. Because CO is much easier to excite than \hh\ in low-temperature environments \citep{1974ApJ...189..441G,1998ApJS..115..241H,2013ARA&A..51..207B},  it is frequently used as a proxy of \hh\ to study dynamical \citep[e.g.,][]{1984ApJ...284..176S, 2005ApJ...625..864Z,  2019ApJS..242...19L}, physical \citep[e.g.,][]{1971ApJ...163L..53S, 2017ApJ...834...57M, 2019MNRAS.488.2970G,2019MNRAS.483.4291C}, and chemical \citep[e.g.,][]{1988ApJ...334..771V, 2009A&A...503..323V, 2019ApJS..243...25W} properties  of molecular clouds. Modern CO surveys are now characterized by large sky coverages, high sensitivities, high spatial and velocity resolutions, and multiple transition and isotopologue lines. \citet{2015ARA&A..53..583H} provide a summary (see Figure 2 therein) of large-scale CO surveys conducted with single-dish telescopes until 2015, and significant progress has been made afterward. To date, the CfA-U.Chile CO survey \citep{2001ApJ...547..792D} has the largest sky coverage ($0\deg <l\leq360\deg$) with an angular resolution of 8.5\arcmin, and a radial velocity resolution of about 0.65 \kms. The spectrum noise is about 0.25 K, which changes slightly from region to region due to different observation strategies.  A new CO survey toward the northern sky ($-10\deg <l\leq250\deg$ and $|b|<5\deg$), the Milky Way Imaging Scroll Painting (MWISP) survey \citep{2019ApJS..240....9S}, records three CO isotopologue lines, \cof, \cos, and \cot, with angular and velocity resolutions of $\sim$50\arcsec\ and $\sim$0.2 \kms, respectively, and an approximate \cofs\ noise root mean squire (rms) of 0.5 K ($\sim$0.3 K for \coss\ and \cots). The FOREST unbiased Galactic plane imaging survey with the Nobeyama 45-m telescope (FUGIN) project \citep{2017PASJ...69...78U} also observes three CO isotopologue lines with a finer angular resolution (20\arcsec) but is less sensitive (1.47 K at a velocity resolution of 1.3 \kms). As for other CO transition lines, the 15-m James Clerk Maxwell Telescope has performed two surveys in the first Galactic quadrant: the \cofs\ ($J=3\rightarrow2$) High-Resolution Survey of the Galactic plane \citep[COHRS,][]{2013ApJS..209....8D} and the \coss/\cots\ ($J=3\rightarrow2$) Heterodyne Inner Milky Way Plane Survey \citep[CHIMPS,][]{2016MNRAS.456.2885R}, with angular resolutions of 14\arcsec\ and 15\arcsec, respectively. Another recent CO survey, covering $^{13}\mathrm{CO}~(J=2\rightarrow1)$ and $\mathrm{C}^{18}\mathrm{O}~(J=2\rightarrow1)$ with the Atacama Pathfinder EXperiment (APEX), is the structure, excitation, and dynamics of the inner Galactic interstellar medium (SEDIGISM) survey \citep[$-60\deg \leq l\leq18\deg$,][]{2017A&A...601A.124S}. 


With the progress of CO surveys, high-dynamic-range statistical analyses of molecular clouds are now achievable. Usually this is done by decomposing large data cubes into small components with appropriate algorithms, such as dendrograms \citep{2008ApJ...679.1338R}, the Spectral Clustering for Interstellar Molecular Emission Segmentation \citep[SCIMES, ][]{2015MNRAS.454.2067C}, FellWalker \citep{2019A&A...632A..58R}, and GaussPy+ \citep{2019A&A...628A..78R}. The definition of molecular clouds, however, is not consistent across different algorithms, and the choice of algorithm hinges on the  desired molecular cloud properties and  specific research goals. For example, GaussPy+ is useful in dynamical  analysis  of molecular clouds \citep{2020A&A...633A..14R}, and the dendrogram and SCIMES are more appropriate in splitting large consecutive structures in position-position-velocity (PPV) space into moderate-sized structures \citep{2019MNRAS.483.4291C}. 


Previous studies have provided many insights into the molecular cloud properties, and in this work, we investigate molecular clouds from another point of view.  The region we analyzed is the local molecular clouds in the first Galactic quadrant ($25.8\deg <l<49.7\deg$, $|b|<5\deg$, and $-6 <V_{\rm LSR}< 30$ \kms) using the CO spectra obtained with the MWISP CO survey. \citet{2019ApJS..240....9S} provided an overview of the MWISP project and molecular clouds in this region in a much wider velocity range ($-80 <V_{\rm LSR}< 130 $ \kms), but we only focus on the local components, which are better resolved. The main component of the local molecular clouds in this region is known as Aquila Rift, or Serpens-Aquila Rift \citep[e.g.,][]{2001ApJ...547..792D,2010A&A...518L..85B,2008ApJ...673L.151G,2020ApJ...893...91S}. We use DBSCAN \citep[density-based spatial clustering of applications with noise,][]{Ester:1996:DAD:3001460.3001507} to extract local molecular cloud samples and examine their statistical properties.



In addition, we use the background-eliminated extinction-parallax (BEEP) method \citep{2019ApJ...885...19Y} to estimate distances to molecular clouds using the \textit{Gaia} DR2 catalog \citep{2016A&A...595A...1G, 2018A&A...616A...1G} supplemented by \av\ estimates \citep{2019A&A...628A..94A}. The BEEP approach was proposed to measure distances to molecular clouds in the Galactic plane where dust environments are complicated. Briefly, the BEEP method removes unrelated extinction by calibrating the stellar extinction toward molecular clouds with the extinction of stars around them, which efficiently reveals the extinction jump position caused by molecular clouds, thus deriving their distances. In the first Galactic quadrant, the dust environment is much more complicated than that in the Outer Galaxy \citep{2019ApJ...885...19Y}, and additional treatments are needed.


We begin the next section (\S\ref{sec:data}) with a description of the CO data, cloud identification methods, and distance calculations. \S\ref{sec:result} describes the distance and statistical results. Discussions about the results are presented in \S\ref{sec:discuss}, and we summarize the conclusions in \S\ref{sec:summary}. 




 \begin{figure*}[ht!]
 \gridline{\fig{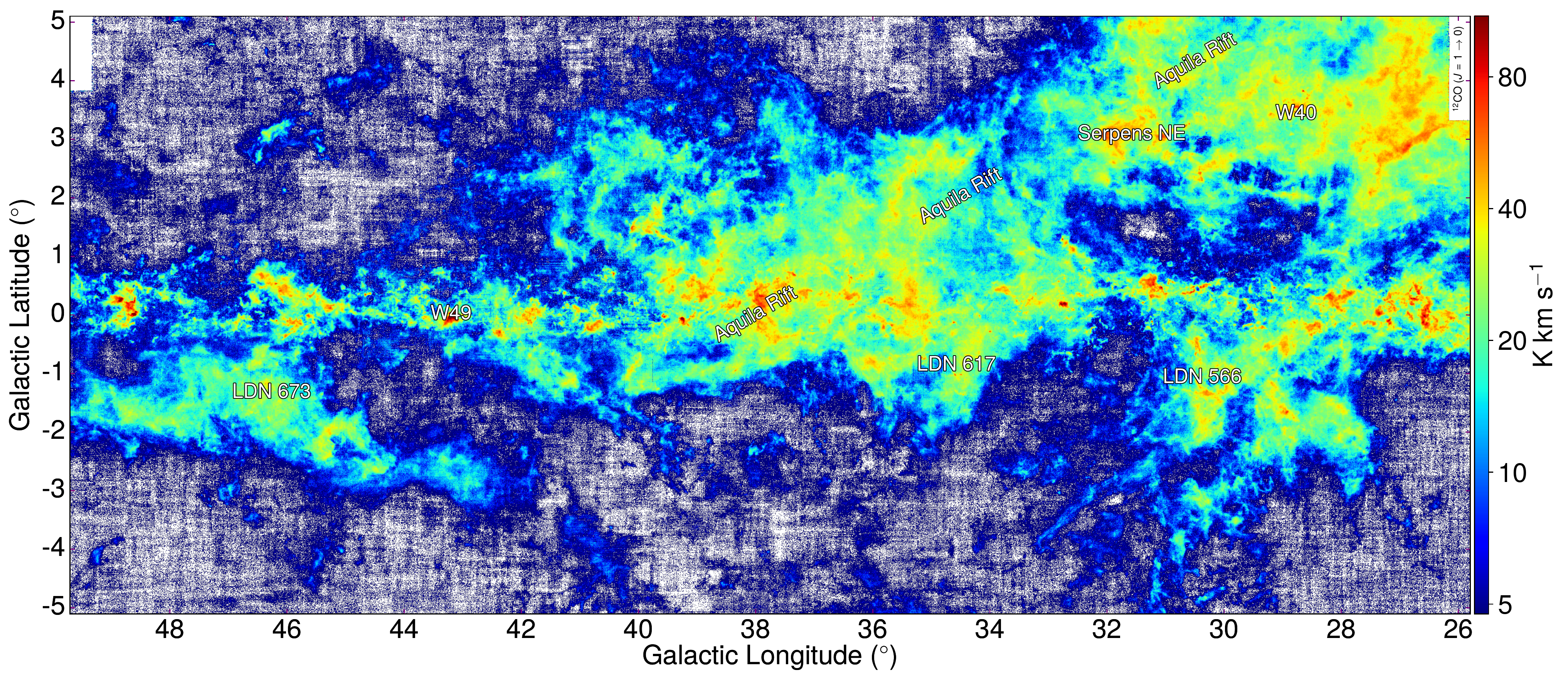}{0.95\textwidth}{(a)} }
 \gridline{\fig{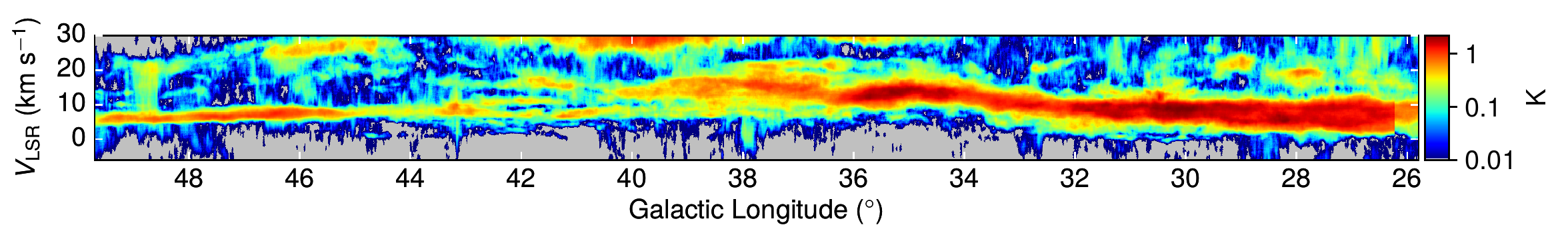}{0.97\textwidth}{(b)}   }
\caption{The integrated intensity map (a)  and the Galactic longitude-velocity ($l$-$V$) diagram (b) of \cof. The radial velocity integration range of the intensity map is [$-6$, 30] \kms\ (the image noise rms is about 1.9 K \kms), and the $l$-$V$ diagram displays the brightness temperature averaged over the Galactic latitudes. Noise voxels in the $l$-$V$ diagram  are masked with DBSCAN (connectivity 1 and MinPts 4) and the post selection criteria down to the 1 K (2$\sigma$) emission level. Distances to prominent regions: W40 \citep[436 pc, ][]{2017ApJ...834..143O}, Serpens NE \citep[465 pc, ][]{2019ApJ...878..111H},    LDN566 \citep[590 pc, ][]{2019ApJ...878..111H}, and W49 \citep[11.11 kpc, ][]{2013ApJ...775...79Z}.
\label{fig:ppvg2650} } 
\end{figure*}






\section{Data And Methods} 
\label{sec:data}

\subsection{CO data} 

The first Galactic quadrant ($25.8\deg <l<49.7\deg$, $|b|<5\deg$, and $-80 <V_{\rm LSR}< 130$ \kms) is a pilot region of the MWISP\footnote{\href {http://www.radioast.nsdc.cn/mwisp.php}{http://www.radioast.nsdc.cn/mwisp.php}} CO survey. The MWISP CO survey covers three CO isotopologue lines, but we only use \cof\ to study the statistical properties of local  components ($-6 <V_{\rm LSR}< 30$ \kms). The reason is that \cofs\ has the lowest critical density and the highest signal-to-noise ratio and produces the most complete molecular cloud  catalog. \citet{2019ApJS..240....9S} have provided a thorough introduction about the MWISP project and an overview of this pilot region, and here we only briefly describe the quality of the data.

Observations were obtained with the Purple Mountain Observatory (PMO) 13.7-m millimeter telescope, and the beam HPBW (Half Power Beam Width) is $\sim$50\arcsec\ for \cofs. The sky region was divided into $30'\times 30'$ tiles, observations of which were merged into a mosaic of the entire CO map. The pixel size of the data cube is $30''\times 30''$, and the velocity channels were regridded to 0.2 \kms. The typical noise rms ($\sigma$) of the  \cofs\ line is $\sim$0.5 K \citep[see Figure 3 of][]{2019ApJS..240....9S}, which changes slightly from tile to tile.

Figure \ref{fig:ppvg2650} shows the integrated intensity map (a) and the $l$-$V$ diagram (b) of the entire region. The radial velocity range of the Local arm is about [$-6$, 25] \kms, and to make the sample complete, we extended the velocity range to [$-6$, 30] \kms, which is the integration range of the  intensity map. In order to make the noise evident, we used a white-color background, and the stripes trace the scan direction. In the $l$-$V$ diagram, the noise was masked with DBSCAN (connectivity 1 and MinPts 4) down to the 1 K (2$\sigma$) emission level, and noise clusters are removed with the post selection criteria. Explanations of DBSCAN and the meaning of its parameters are presented in \S\ref{sec:cloudIdentification}.

 \begin{figure*}[ht!]
 \gridline{\fig{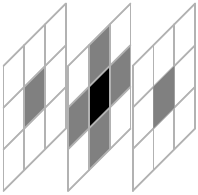}{0.3\textwidth}{(a) connectivity 1 }  \fig{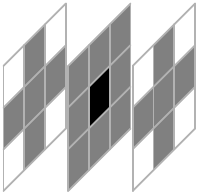}{0.3\textwidth}{(b) connectivity 2}   \fig{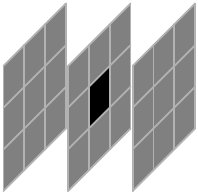}{0.3\textwidth}{(c) connectivity 3} 
 }

\caption{Three types of connectivity. Gray points are neighbors of the black point in each connectivity type. \label{fig:contype} } 
\end{figure*}

\subsection{Cloud detection}
 \label{sec:cloudIdentification}


 In this study, we examine molecular clouds in general, and ignore their hierarchical details, i.e., molecular clouds are defined as consecutive structures in PPV space. Statistical results are based on properties of those independent PPV structures. With this definition, we use DBSCAN\footnote{\href {https://scikit-learn.org/stable/modules/generated/sklearn.cluster.DBSCAN.html}{https://scikit-learn.org/stable/modules/generated/sklearn.cluster.DBSCAN.html}} to identify clusters as molecular clouds. Trunks of the  dendrogram\footnote{\href {https://dendrograms.readthedocs.io/en/stable}{https://dendrograms.readthedocs.io/en/stable}} also satisfy these requirements, but if only trunks were concerned, the dendrogram would be a special case of DBSCAN. Consequently, we do not use the dendrogram, but explore DBSCAN parameters and demonstrate the variation of statistical results. 

 In addition to DBSCAN, we also use SCIMES\footnote{\href {https://scimes.readthedocs.io/en/latest}{https://scimes.readthedocs.io/en/latest}} to decompose large molecular clouds, but statistics with SCIMES are only used for comparison. Hierarchical Density-Based Spatial Clustering of Applications with Noise (HDBSCAN), an improved version of DBSCAN, is also able to identify clusters, and as discussed in \S\ref{sec:hdbscan}, HDBSCAN does not provide a uniform molecular cloud definition in PPV space. Therefore, we did no use HDBSCAN.

 Two parameters of DBSCAN, $\epsilon$ and MinPts, define the connection property of voxels in PPV space.   The number of points in an $\epsilon$ radius is assigned as the density, and a minimum density of MinPts is required to be considered as core points. Directly connected (within an $\epsilon$ radius) core points form clusters, and neighbors (within an $\epsilon$ radius) of those core points are also considered as their cluster members. Figure \ref{fig:contype} shows three types of connectivity in PPV space,  i.e., definitions of neighborhood. Connectivity 1  corresponds to $\epsilon$ = 1 in Euclidean distance  (6 neighboring points at most), $\epsilon=\sqrt{2}$ (18 points) for connectivity 2  and  $\epsilon=\sqrt{3}$ (26 points) for connectivity 3. If only trunks were considered, the dendrogram would correspond to DBSCAN with $\epsilon = 1$ and MinPts = 3.  We examined all three types of connectivity, and the choice of the minimum MinPts for each connectivity type is demonstrated in \S\ref{sec:minpts}.


 We examine the effects of CO emission cutoffs on statistical results. The cutoff ranges from 2$\sigma$ (1 K) to 7$\sigma$ with a step of 0.5$\sigma$, and before running DBSCAN, all voxels below cutoffs are removed.

 The definition of molecular clouds in SCIMES is not straightforward, due to its multistage procedure and uncertain mathematical properties. SCIMES extracts clusters from the dendrogram results, and the dendrogram is a single-linkage hierarchical clustering algorithm, which produces tree structures according to the intensities and Euclidean  distances of voxel indices. Consequently, the SCIMES results are only used for a comparison purpose in statistical analyses, but SCIMES is able to split large dendrogram trunks into medium-sized branches, which is useful in distance examinations. The dendrogram builds tree structures from the highest intensities (leaves) down to the cutoff value (roots), and in the tree structure, branches are structures that contains leaves or branches. The dendrogram has three parameters, min\_value, min\_delta, and min\_npix. Min\_value is a cutoff threshold of the intensity, i.e., all voxels below min\_value are  discarded. Min\_npix is the minimum number of voxels that a cloud has to have,  and min\_delta is the minimum difference between the peak and the background. The background is defined as the value that leaves merge into parent branches. For isolated leaves, the background is min\_value, so their minimum peak values are (min\_value+min\_delta). To see the effect of parameters,  we fixed min\_delta (3$\sigma$) and min\_npix (16), and changed min\_value (from 2$\sigma$ to 7$\sigma$ with a step of 0.5$\sigma$), examining 11 cases in total. SCIMES takes the dendrogram tree structure as a fully connected graph (in a single trunk) and splits large trunks based on the magnitude of their volume and luminosity, or other self-defined criteria \citep{2015MNRAS.454.2067C}.    In SCIMES, we saved all structures, including isolated and unclustered leaves or branches.


 \begin{figure*}[ht!]
 \plotone{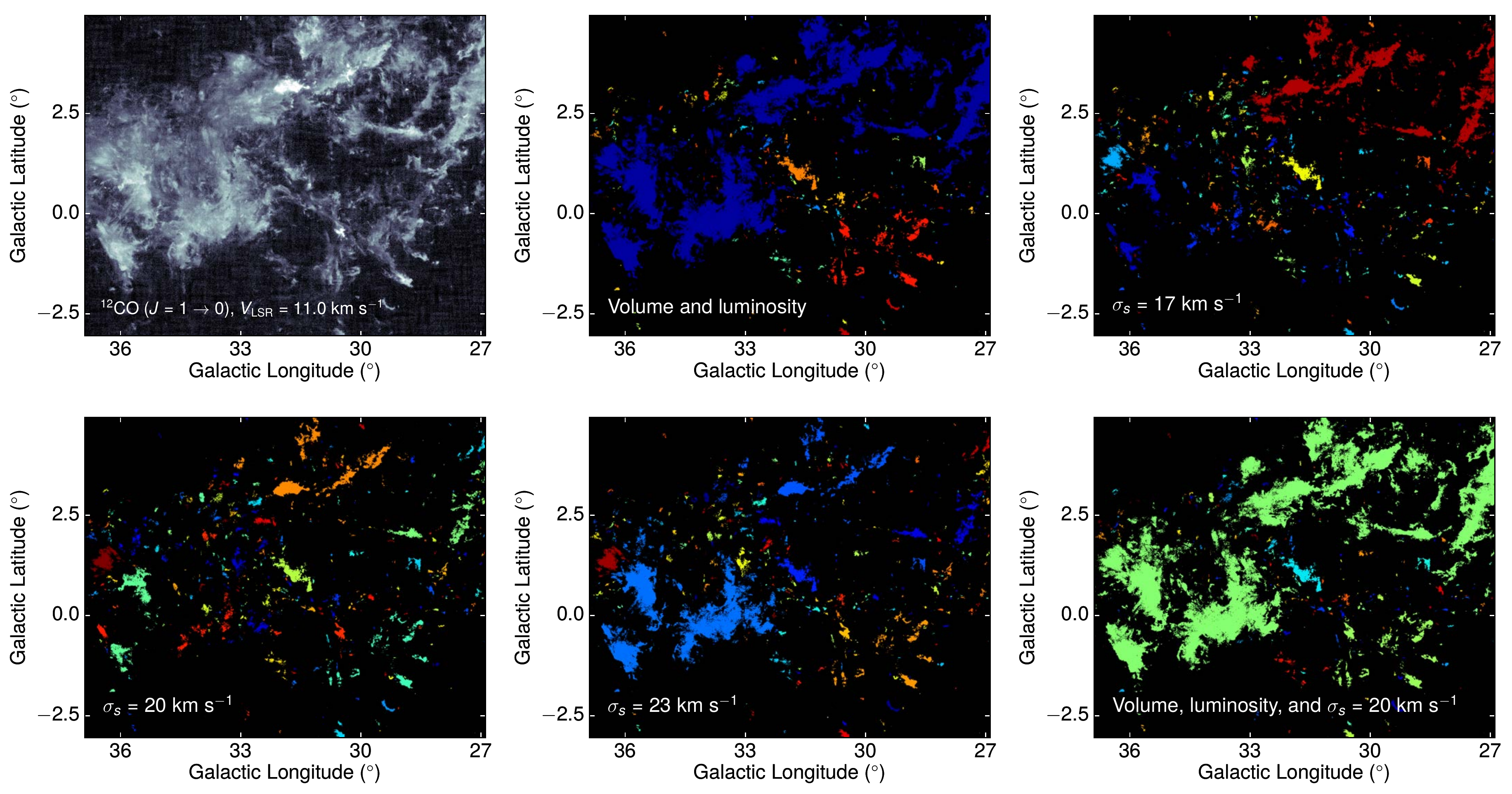}
\caption{SCIMES results with different criteria. The CO cutoff is 1.5 K (3$\sigma$),  min\_npix = 16, and min\_delta = 1.5 K. See Equation \ref{eq:sigmas} for the meaning of $\sigma_s$. \label{fig:scimescriteria}}
\end{figure*}


We found that the default volume and luminosity criteria did not work well in the Aquila-Rift region, so we used the radial velocity range as the criterion to split large trunks with SCIMES. The reason is that far molecular clouds usually have small volume and low luminosity compared with local molecular clouds, and far molecular clouds would be treated as small molecular clouds if a constant volume and luminosity criteria were used. The radial velocity, however, is independent of the distance. The second moment of radial velocities overwhelms many weak branches, so we adopted instead the minimum velocity range that contains all leaves in a branch as a proxy of its velocity dispersion. For isolated leaves, the second moment of the radial velocity is still used. The velocity dispersion is rescaled with $\sigma_s$, and because SCIMES results are only used for comparison, and the choice of $\sigma_s$ is unimportant as long as large trunks are able to be decomposed. The rescaled radial velocity range is 
\begin{equation}
s_{ij}=\exp\left( \frac{  -v_{ij}^2  }{  \sigma_s^2 }  \right),
\label{eq:sigmas}
\end{equation}
where $v_{ij}$ is the radial velocity range of the smallest branch that contains both the $i$th and $j$th structures, which is symmetric. 


In Figure \ref{fig:scimescriteria}, we compare SCIMES results with different splitting criteria. Unlike the radial velocity range, the default volume and luminosity criteria kept a large portion of the largest dendrogram trunks in a whole piece. Because it is difficult to determine which criteria are the best and the SCIMES results are only used to see how the statistical results vary if large dendrogram trunks were decomposed, we simply used a moderate value of 20 \kms. The choice of criteria is rather subjective, and this is one of the reasons that SCIMES results are only used for comparison.



\subsection{Post Selection Criteria}

Clusters identified by DBSCAN may contain noises, so we apply post selection criteria on raw clusters based on the voxel number, the peak brightness temperature, the area projection, and the velocity projection. The first two criteria are related to sensitivity, while the second two are related to resolution. In total, the post selection criteria contain four conditions: (1) the minimum voxel number is 16; (2) peak intensity $\geq$ (min\_value+3$\sigma$); (4) the projection area contains a beam, i.e., a  compact 2$\times$2 region; (4) velocity channel numbers $\geq$ 3. We examined all criteria and used them in combination, and these criteria also apply to SCIMES clusters.

\begin{figure*}[ht!]
 \gridline{\fig{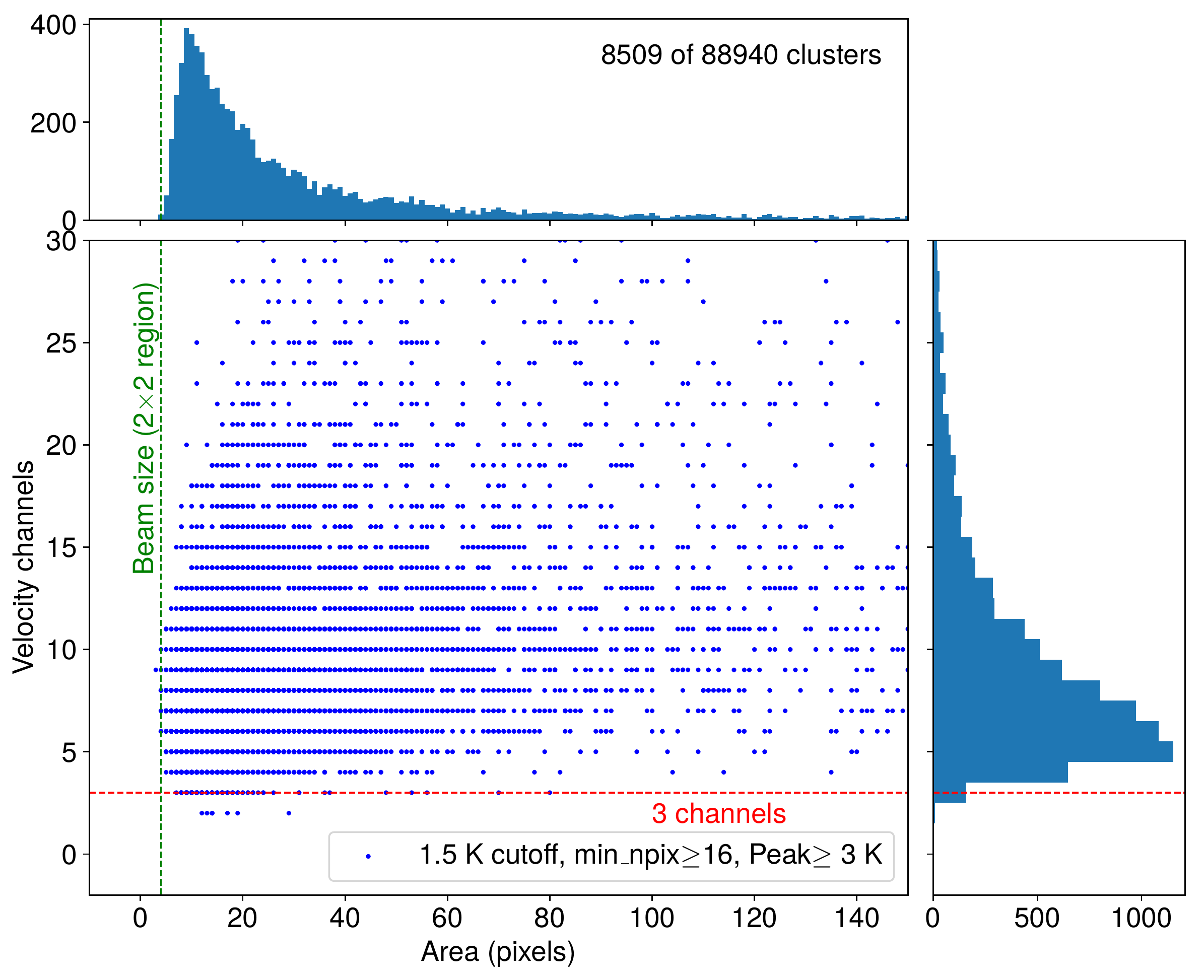}{0.45\textwidth}{(a)}  \fig{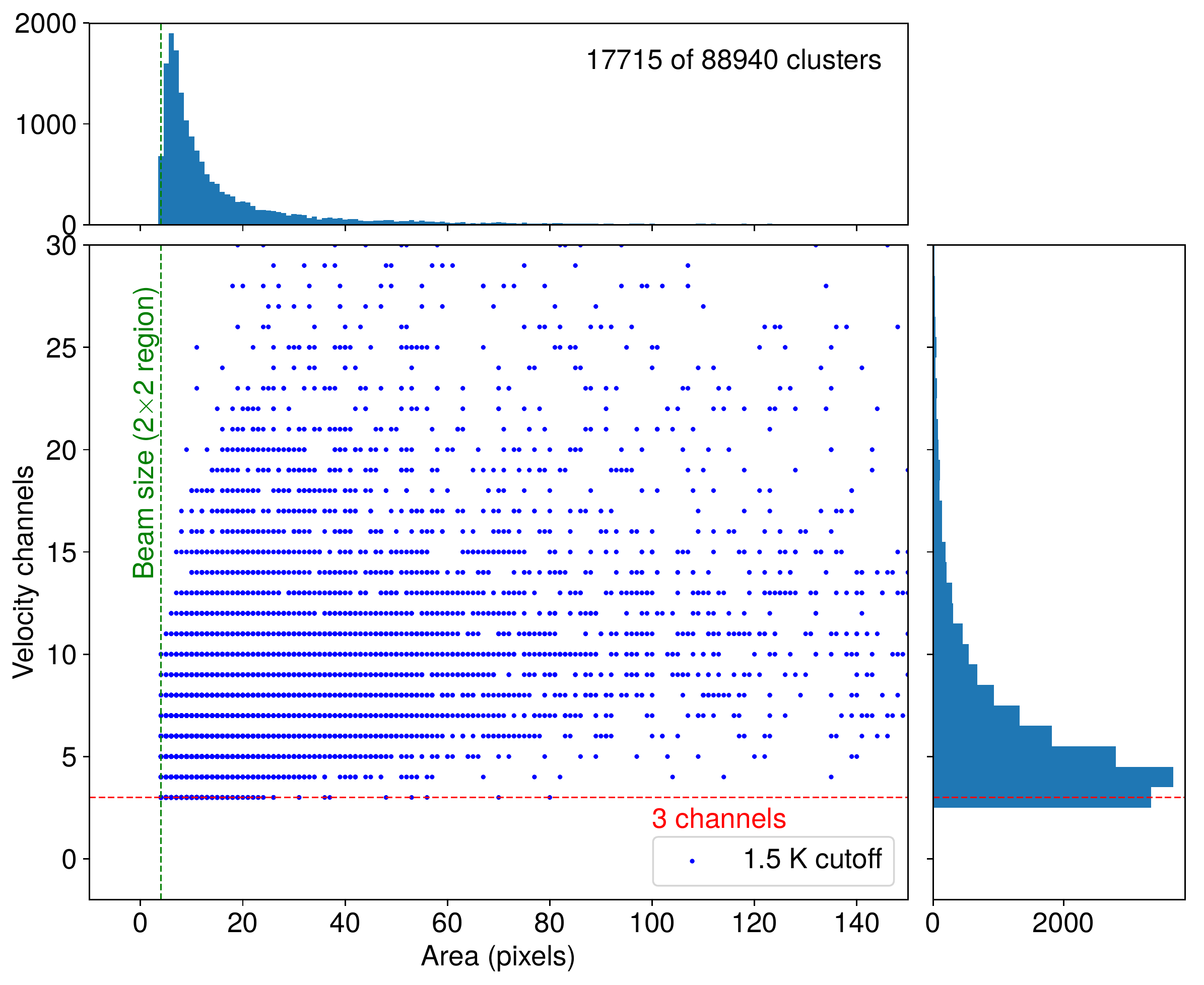}{0.45\textwidth}{(b)}  }
\caption{Post selection criteria: select by (a) minimum voxel numbers and peak intensities and by (b) minimum area and velocity projection. Raw samples are produced by DBSCAN with the 1.5 K (3$\sigma$) cutoff, connectivity 1, and MinPts 4. A beam means a compact 2$\times$2 region. \label{fig:criteria} }
\end{figure*}

Figure \ref{fig:criteria} describes the effect of the sensitivity and resolution criteria separately. The histogram shows that most clusters selected by sensitivity satisfy the resolution criteria, indicating that the sensitivity criteria is more strict. In practice, we used both criteria in combination, but the sensitivity criteria dominate the post selection.

\subsection{Choice of MinPts}
\label{sec:minpts}

Small values of MinPts include many noises, even after applying the post selection criteria,  and we use negative brightness temperatures in spectra to demonstrate the choice of the minimum MinPts. Obviously, negative values are all noises, and after selecting with the post criteria, no DBSCAN clusters should remain.

\begin{figure*}[ht!]
\plotone{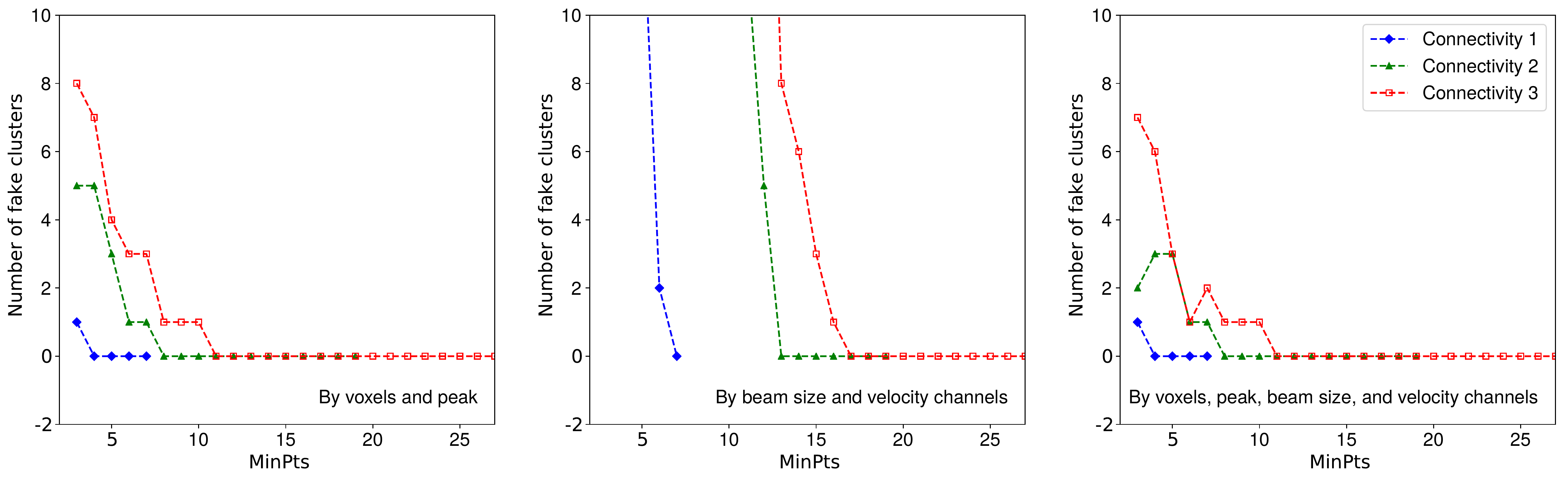}
\caption{Cluster numbers detected with negative CO brightness temperatures. \label{fig:negative} }
\end{figure*}

We inverse the sign of spectra, and impose a 1 K (2$\sigma$) cutoff. All possible cases of DBSCAN were examined based on this data set, and the number of remaining clusters after applying the post selection criteria is shown in Figure \ref{fig:negative}. The minimum MinPts values that result in 0 clusters for connectivity 1, 2, and 3 are 4, 8, and 11, respectively, and reasonably, larger $\epsilon$ requires higher MinPts.

\begin{figure*}[ht!]
 
\gridline{\fig{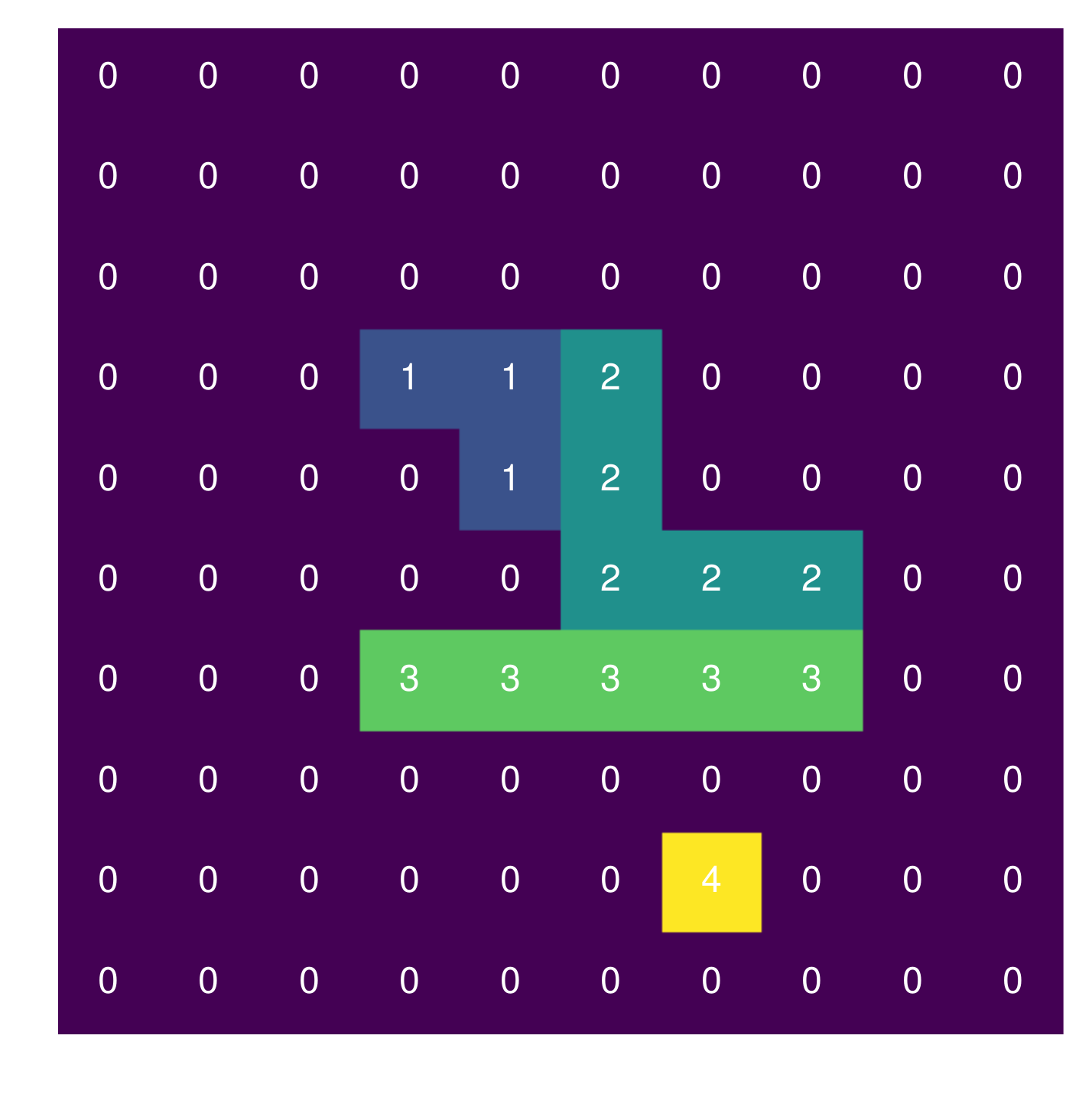}{0.45\textwidth}{(a) Before dilation} \fig{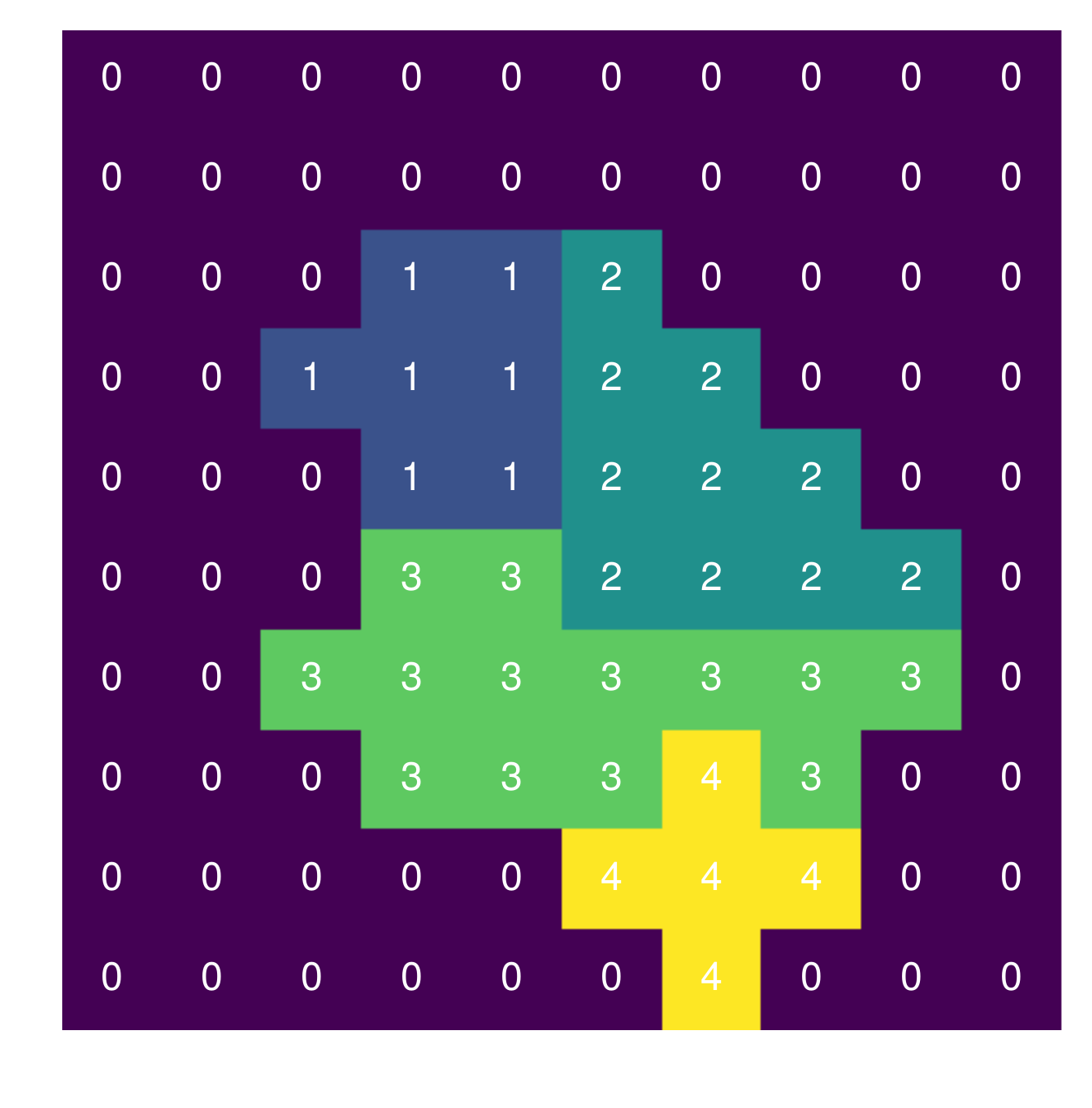}{0.45\textwidth}{(b) After dilation}  }
\caption{Regions (a) before and (b) after one step of two-dimensional dilation. In the dilation, larger label numbers are preferred when multiple clusters are present in the neighbor of unlabeled pixels. After dilation, the area of regions becomes larger. \label{fig:dilation} }
\end{figure*}

\subsection{Distances}
\label{sec:discloud}

We examined distances to large molecular clouds with the  BEEP method proposed by \citet{2019ApJ...885...19Y}. The BEEP method removes irregular variations of stellar  extinction that are unrelated to molecular clouds, and thus reveals the extinction jump point caused by targeted  molecular clouds.

We used parallaxes and \ag\ in the \textit{Gaia} DR2 catalog,  supplemented by \av\ \citep{2019A&A...628A..94A} when \ag\ stars were insufficient.  \citet{2019A&A...628A..94A} calibrated systematic errors (see Table 1 therein) of \textit{Gaia} DR2 data, and in order to make the results consistent, we performed the same calibration. As suggested by \citet{2019A&A...628A..94A}, we require all flags of the \av\ values to be 0, i.e., those data are reliable. Other procedures of \textit{Gaia} DR2 data reductions follow  \citet{2019ApJ...885...19Y}.


Because distance calculations favor moderate-sized molecular clouds, we used the dendrogram and SCIMES to decompose the data cube. The area of molecular clouds needs to be adequate to include sufficient on-cloud stars, and the flux needs to be large enough to cause detectable extinction. In the approximate 239 deg$^2$ area and within 1 kpc, the total number of stars that have both good parallaxes (relative error $<$ 0.2) and \av\ values (all flags are 0) is 320787, and the number of stars per unit area is about 1342. To get a 10\% distance precision at 500 pc, we require that there is at least 1 star per 50 pc, so the total number of on-cloud stars within 1 kpc should be at least 20, corresponding to an area of 0.015 deg$^2$ (216 pixels). Considering on-cloud stars need to be located in regions that show significant CO emission, a value of 1000 for min\_npix (min\_value = 3$\sigma$ and min\_delta = 3$\sigma$) is small enough to generate a complete catalog for distance measurements. With these parameter settings, we run the dendrogram, and the criteria used in SCIMES is the velocity range (see \S\ref{sec:cloudIdentification}, $\sigma_s$ = 20 \kms) instead of the volume and luminosity criteria. In total, SCIMES found 912 clusters, including unclustered structures.

SCIMES only selects substructures from large trunks, and misses weak emission in the envelopes of molecular clouds.  The reason is that the dendrogram uses high contour levels to distinguish substructures, and emission  with lower contour levels is discarded if large trunks were replaced with their branches. To add back the weak emission, we use the \texttt{dilation} function in  the \texttt{Python} \texttt{skimage} package to expand labeled SCIMES clusters to 2$\sigma$. By default, \texttt{dilation} extends large regions and shrinks small regions, so we restricted \texttt{dilation} only to expand into unlabeled regions. An example of one dilation step is shown in Figure \ref{fig:dilation}, and after the dilation, unassigned 2$\sigma$ cutoff regions are assigned with the largest neighboring (connectivity 1) labels. This type of dilation prefers large label numbers near cluster boundaries, and because it is hard to determine which clusters those edge voxels should belong to, we simply follow this rule of \texttt{dilation}. Because \texttt{dilation} ignores CO values and to make sure molecular clouds split at valleys, we expanded labeled regions gradually by decreasing contour levels from 10$\sigma$ to 2$\sigma$ with a step of 1$\sigma$. At each step, the \texttt{dilation} runs recursively until all unlabeled voxels above the targeted contour level are assigned to adjacent SCIMES clusters. The dilation process is similar to expanding mountain peaks to mountain roots, and mountains are separated at valleys.

Compared with molecular clouds in the Outer Galaxy \citep{2019ApJ...885...19Y}, the choice of on- and off-cloud regions in the first Galactic quadrant is slightly different, and stars that are effected by foreground clouds were removed. Given the complexity of molecular cloud environments, on- and off-cloud regions should neither be too large nor  too small. Large regions contain contamination from other molecular clouds, making the baseline unreliable,  while small regions contain insufficient stars, unable to yield robust statistical results. Consequently, we extended the region box (along $l$ and $b$) that contains molecular clouds by 0.5\deg. If a molecular cloud is both large and connected with many adjacent ones, we manually chose a on-cloud region where the cloud has sharp boundaries. If a molecular cloud has not enough on-cloud stars and is likely to be associated with adjacent emissions, we extended the on-cloud region to include more on-cloud stars.

We removed those stars (both in on- and off-cloud stars) that are affected by foreground clouds. The foreground emission of a molecular cloud is defined by means of its  weighted (by the brightness temperature) mean ($V_{\rm center}$) and weighted standard deviation  ($\Delta_V$) of the radial velocity.  Integrated intensity from $-6$ \kms\ to  ($V_{\rm center}-3\Delta_V$) are taken as foreground molecular cloud emission.  Stars, toward which the foreground emission is larger than 3 K \kms, were removed. Although it is possible that some clouds are farther than the targeted molecular cloud, removing those stars would only make the background baseline more reliable, which does no harm to the distance calculation. This step would largely guarantee that the first jump point along the line of sight is due to the targeted molecular cloud.

The upper threshold of CO emission toward off-cloud stars (the noise level) is set as 1.5 K \kms\ ($\sim$1.5$\sigma$), above which off-cloud stars were removed, while the lower threshold of CO emission toward on-cloud stars (the signal level) was 3-5 K \kms, below which on-cloud stars were removed. In some cases, lower signal levels (3 K \kms) were used to include more on-cloud stars to obtain robust statistical results.

In total, we examined 400 molecular clouds, and  derived distances to 28 of them. Five reasons are responsible for unsuccessful  distance calculations.  First, many molecular clouds are too weak to cause detectable \av\ or \ag.  Secondly, some near molecular clouds have too few foreground stars, and their distances cannot be firmly constrained. Thirdly,  a number of molecular clouds have no clear boundaries, and no nearby off-cloud regions are available. Those clouds are possibly associated with adjacent components. Fourthly, many molecular clouds have heavy foreground emission, and both the \av\ and \ag\ data are truncated. Fifthly, there are molecular clouds that are too far ($>$ 2.5 kpc) to be local components, and no separation of on- and off-cloud extinction were seen. We started from molecular clouds that have larger angular areas and stopped  calculating when consecutively 100 molecular clouds have no distance detection. Results of the 28 molecular clouds are presented in \S\ref{sec:distance}.

%


\section{Results}
\label{sec:result}

This section presents the distances and statistical properties of molecular clouds. Statistics include the total number of molecular cloud samples, the equivalent linewidth, the peak brightness temperature, the physical area, and the mass. The variation of statistical results with different molecular cloud definitions is gauged by changing cutoffs, connectivity types, and MinPts values. Generally, we have three types of molecular cloud samples: (1) relatively large molecular clouds only used for distance examination produced with SCIMES (extended to 2$\sigma$ level);  (2) samples with small molecular clouds produced with SCIMES; (3) samples produced with DBSCAN. Statistics are based on DBSCAN samples, the second type of samples are only used for comparison.  



\subsection{Distances}
\label{sec:distance}

\begin{figure*}[ht!]
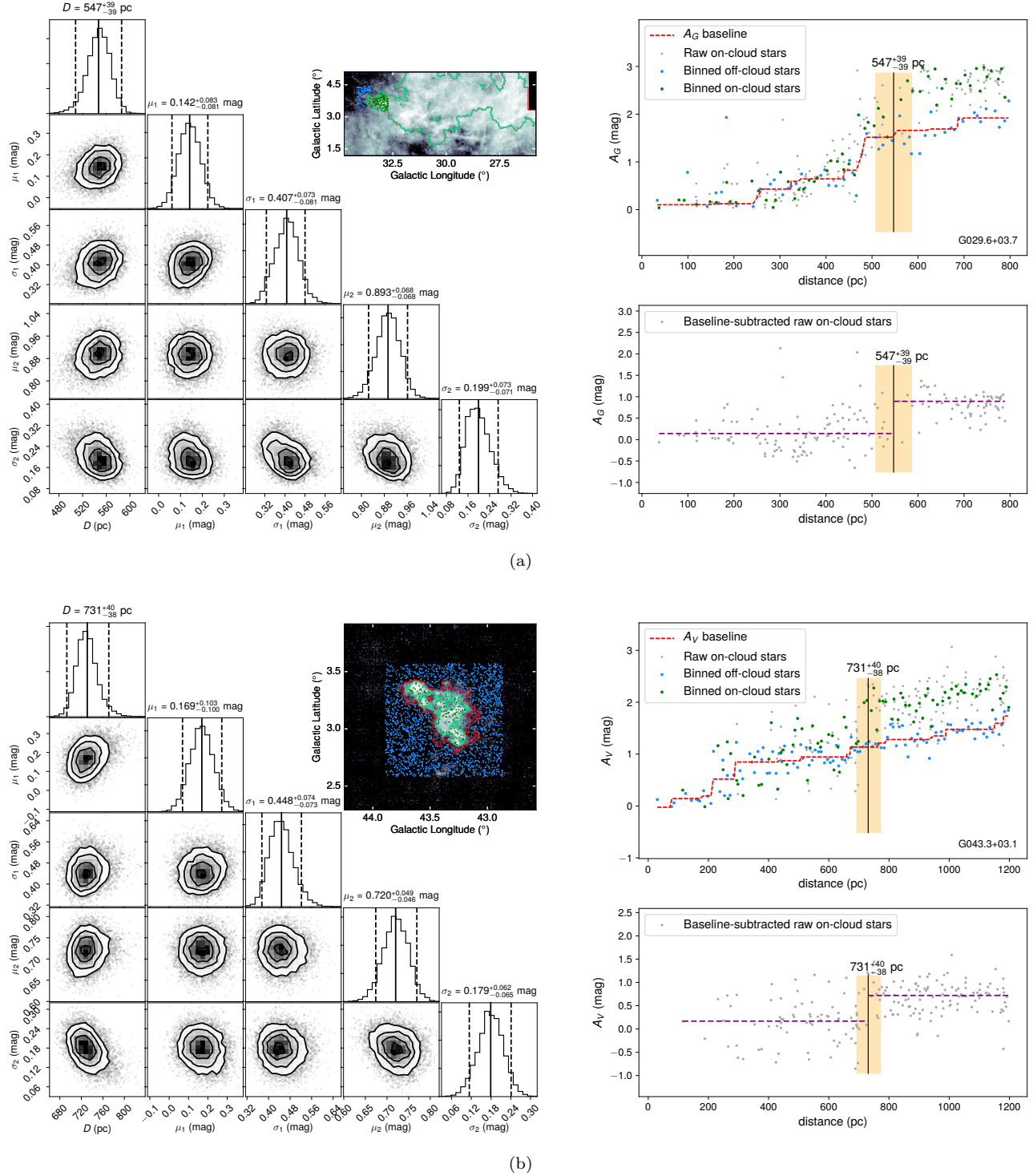

 \gridline{\fig{{G029.6+03.7DistanceAG}.pdf}{0.95\textwidth}{(a)} }
 \gridline{\fig{{G043.3+03.1DistanceAV}.pdf}{0.95\textwidth}{(b)} }
\caption{Distances to two molecular clouds, G029.6+03.7 with \ag\ (a) and G043.3+03.1 with \av\ (b). On the right-hand side plots of both panels, the red line is the baseline obtained with off-cloud stars, and the broken horizontal purple line is the extinction variation of on-cloud stars after subtracting the baseline. In the cloud CO image insets, the green and light blue points represent samples of on- and off-cloud stars, respectively. The red lines delineate the area range of molecular clouds, while the green lines are the lower threshold of CO emission towards on-cloud stars. In the image insets of panel (a), the off-cloud region is selected to be close to the on-cloud region to accurately remove irregular extinction variations of on-cloud stars. The left corner maps are the MCMC sampling results of five  parameters: the distance ($D$), the mean and standard deviation of foreground extinction ($\mu_1$ and $\sigma_1$), and the mean and standard deviation of background extinction ($\mu_2$ and $\sigma_2$). See \citet{2019ApJ...885...19Y} for other details.  \label{fig:twoclouds} }
\end{figure*}

We present distance results for 28 local molecular clouds in the first Galactic quadrant. \ag\ is used for 20 molecular clouds, while  the rest eight were derived with \av\ \citep{2018A&A...616A...8A}.  For both \av\ and \ag, we require that the jump point is evident and the background on-cloud stars are clearly above the baseline. As examples, in Figure \ref{fig:twoclouds}, we demonstrate distances to two molecular clouds, G029.6+03.7 (\ag) and  G043.3+03.1 (\av).



\begin{deluxetable*}{ccccccccccccc}
\tabletypesize{\scriptsize}
\tablecaption{Distances to 28 molecular clouds.\label{Tab:cloudDis}}
\tablehead{
   Name &  $l$  &   $b$   &\colhead{$V_{\rm LSR}$} &  Area &  $D_{\rm Gaia}$\tablenotemark{a} &   N\tablenotemark{b}   & CO$_{\rm cut}$\tablenotemark{c}   &  $D_{\rm cut}$   &  Mass\tablenotemark{d}      & $D_{\rm kinematic}$\tablenotemark{e} &Extinction & Note \\
   &  (\deg)  &  (\deg)  & (\kms)  &  deg$^2$ &   (pc)   &  & (K)   & \colhead{(pc)}&   ($10^3$ \msun)     &   (kpc)     
}
\colnumbers
\startdata
 G026.9$-$03.5 &    26.980 &  -3.528 &    16.3  &  0.15 & $ 425_{-  57}^{+  38}$ &    80 &  3 &  800  &   0.1 & $1.16_{-0.51}^{+ 0.47}$ & $A_V$ &         \\  
 G027.8$-$02.1 &    27.891 &  -2.172 &    18.0  &  1.85 & $ 487_{-  18}^{+  15}$ &   407 &  5 & 1000  &   4.6 & $1.26_{-0.49}^{+ 0.45}$ & $A_G$ &         \\  
 G028.8$-$01.9 &    28.894 &  -1.921 &     3.9  &  4.20 & $ 261_{-   9}^{+   9}$ &   362 &  5 &  500  &   3.1 & $0.21_{-0.16}^{+ 0.53}$ & $A_G$ &         \\  
 G029.6$+$03.7 &    29.635 &   3.739 &     7.4  & 15.45 & $ 547_{-  39}^{+  39}$ &   170 &  3 &  800  & 114.9 & $0.49_{-0.15}^{+ 0.50}$ & $A_G$ & W40     \\  
 G030.3$+$01.2 &    30.388 &   1.283 &     5.2  &  0.26 & $ 428_{-  54}^{+  47}$ &    53 &  5 & 1000  &   0.2 & $0.32_{-0.16}^{+ 0.51}$ & $A_G$ &         \\  
 G030.6$-$03.6 &    30.627 &  -3.649 &     9.4  &  2.09 & $ 317_{-  39}^{+  58}$ &   105 &  5 &  600  &   1.3 & $0.62_{-0.53}^{+ 0.49}$ & $A_G$ &         \\  
 G035.0$+$00.6 &    35.098 &   0.656 &    12.8  & 21.73 & $ 606_{-   9}^{+   9}$ &  1267 &  5 & 1000  & 130.2 & $0.84_{-0.48}^{+ 0.45}$ & $A_G$ &    Phoenix cloud\tablenotemark{f}     \\  
 G036.6$-$00.9 &    36.671 &  -0.975 &    12.7  &  0.29 & $ 588_{-  58}^{+  67}$ &   142 &  3 & 1200  &   0.4 & $0.83_{-0.48}^{+ 0.45}$ & $A_G$ &         \\  
 G037.0$-$01.4 &    37.061 &  -1.420 &    18.6  &  0.30 & $ 528_{-  46}^{+  39}$ &   118 &  5 &  800  &   0.5 & $1.20_{-0.45}^{+ 0.43}$ & $A_V$ &         \\  
 G037.5$+$04.0 &    37.546 &   4.019 &    15.5  &  0.32 & $ 515_{-  24}^{+  22}$ &   279 &  5 & 1000  &   0.5 & $1.01_{-0.46}^{+ 0.44}$ & $A_V$ &         \\  
 G038.3$-$00.1 &    38.393 &  -0.187 &    16.2  &  7.03 & $ 645_{-  22}^{+  22}$ &   445 &  5 & 1200  &  44.7 & $1.05_{-0.46}^{+ 0.44}$ & $A_G$ &         \\  
 G039.4$-$02.7 &    39.466 &  -2.732 &    14.9  &  0.17 & $ 492_{-  39}^{+  45}$ &   109 &  3 & 1000  &   0.1 & $0.96_{-0.46}^{+ 0.44}$ & $A_V$ &         \\  
 G040.7$-$04.1 &    40.729 &  -4.114 &     7.8  &  2.34 & $ 377_{-   9}^{+  10}$ &   401 &  5 &  700  &   1.5 & $0.49_{-0.49}^{+ 0.46}$ & $A_V$ &         \\  
 G041.3$+$03.2 &    41.343 &   3.214 &    26.9  &  0.17 & $1071_{-  81}^{+  77}$ &   135 &  5 & 1500  &   0.8 & $1.73_{-0.44}^{+ 0.43}$ & $A_G$ &         \\  
 G041.5$+$02.3 &    41.545 &   2.322 &    17.7  &  0.92 & $ 905_{-  50}^{+  48}$ &   111 &  7 & 1500  &   4.4 & $1.15_{-0.46}^{+ 0.44}$ & $A_G$ &         \\  
 G042.0$-$00.9 &    42.036 &  -0.981 &    10.7  &  2.31 & $ 719_{-  28}^{+  27}$ &   254 &  5 & 1000  &   8.8 & $0.69_{-0.48}^{+ 0.46}$ & $A_G$ &         \\  
 G043.2$+$02.1 &    43.201 &   2.141 &    23.1  &  0.31 & $ 872_{-  84}^{+  77}$ &    91 &  4 & 1500  &   0.4 & $1.51_{-0.45}^{+ 0.44}$ & $A_G$ &         \\  
 G043.3$+$03.1 &    43.385 &   3.152 &    10.2  &  0.20 & $ 731_{-  38}^{+  40}$ &   190 &  4 & 1200  &   0.4 & $0.67_{-0.48}^{+ 0.46}$ & $A_V$ &         \\  
 G044.5$+$02.6 &    44.551 &   2.700 &    14.6  &  0.17 & $ 772_{- 138}^{+ 124}$ &    59 &  3 & 1500  &   0.3 & $0.97_{-0.47}^{+ 0.46}$ & $A_G$ &         \\  
 G044.8$+$04.0 &    44.802 &   4.008 &    20.1  &  0.66 & $ 859_{-  26}^{+  29}$ &   319 &  5 & 1500  &   2.1 & $1.33_{-0.46}^{+ 0.46}$ & $A_G$ &         \\  
 G045.5$-$04.3 &    45.538 &  -4.326 &    18.6  &  0.32 & $ 938_{-  60}^{+  61}$ &   139 &  4 & 1500  &   1.0 & $1.24_{-0.47}^{+ 0.46}$ & $A_G$ &         \\  
 G046.2$+$03.1 &    46.207 &   3.119 &    26.9  &  0.32 & $ 937_{-  32}^{+  35}$ &   256 &  5 & 1500  &   2.5 & $1.80_{-0.47}^{+ 0.47}$ & $A_G$ &         \\  
 G046.2$-$01.6 &    46.284 &  -1.660 &     7.1  & 15.37 & $ 394_{-   5}^{+   4}$ &  1907 &  5 &  800  &  22.2 & $0.47_{-0.48}^{+11.08}$ & $A_G$ &         River cloud\tablenotemark{f}   \\ 
 G046.5$-$03.3 &    46.599 &  -3.373 &    27.5  &  0.15 & $1348_{-  98}^{+  98}$ &   225 &  3 & 2000  &   1.1 & $1.84_{-0.47}^{+ 0.48}$ & $A_G$ &         \\  
 G046.9$-$02.9 &    46.991 &  -2.986 &    26.4  &  0.24 & $1300_{-  66}^{+  67}$ &   241 &  5 & 2000  &   2.4 & $1.78_{-0.47}^{+ 0.48}$ & $A_G$ &         \\  
 G048.3$-$01.7 &    48.385 &  -1.713 &    21.4  &  2.40 & $1134_{-  77}^{+  82}$ &   260 &  5 & 2000  &  18.5 & $1.47_{-0.48}^{+ 0.49}$ & $A_G$ &         \\  
 G048.9$+$02.3 &    48.906 &   2.313 &     6.8  &  0.91 & $ 483_{-  32}^{+  29}$ &   233 &  4 &  700  &   0.8 & $0.47_{-0.49}^{+10.51}$ & $A_V$ &         \\  
 G049.0$+$03.4 &    49.053 &   3.431 &     6.4  &  0.22 & $ 480_{-  35}^{+  34}$ &   127 &  4 & 1000  &   0.2 & $0.44_{-0.49}^{+10.53}$ & $A_V$ &         \\  
\enddata
\tablenotetext{a}{The 5\% systematic error is not included, and figures of those 28 molecular clouds are publicly accessible on the Harvard Dataverse (\href{https://doi.org/10.7910/DVN/8HAPXB}{https://doi.org/10.7910/DVN/8HAPXB}). }
\tablenotetext{b}{Total number of on-cloud stars.  } 
\tablenotetext{c}{The lower threshold of CO emission towards on-cloud stars.  } 
\tablenotetext{d}{The mass only takes account of CO-bright molecular gas.  } 
\tablenotetext{e}{Derived from the A5 model of \citet{2014ApJ...783..130R}.} 
\tablenotetext{f}{ ``Phoenix cloud'' and ``River cloud'' (LDN 673) are nicknames provided by \citet{2020ApJ...893...91S}. } 
\end{deluxetable*}

Table \ref{Tab:cloudDis} summarizes the distance results of the 28 molecular clouds. The nearest distance of those molecular clouds is 261 pc, while the farthest one is 1348 pc. From left to right, we display the name (1), averaged $l$-$b$-$V$ position (2, 3, 4), the angular area (5), \textit{Gaia} DR2 distances (6), number of on-cloud stars (7), minimum CO emission of on-cloud stars (8), maximum distances to on-cloud stars  (9), molecular cloud masses (10), kinematic distances (11), the extinction used (12),  and notes (13). The molecular cloud masses are  estimated by assuming a $^{12}$CO-to-H$_2$ mass conversion factor of X = $2.0\times 10^{20}$ cm$^{-2}$ $\rm(K\  km\ s^{-1})^{-1}$ \citep{2013ARA&A..51..207B}, and the kinematic distances are derived with the A5 model of \citet{2014ApJ...783..130R}. The systematic distance error is about 5\% \citep{2019A&A...624A...6Y,2020A&A...633A..51Z}, which is possibly larger for far-distance molecular clouds ($>$ 1 kpc).


\begin{figure*}[ht!]
 \plotone{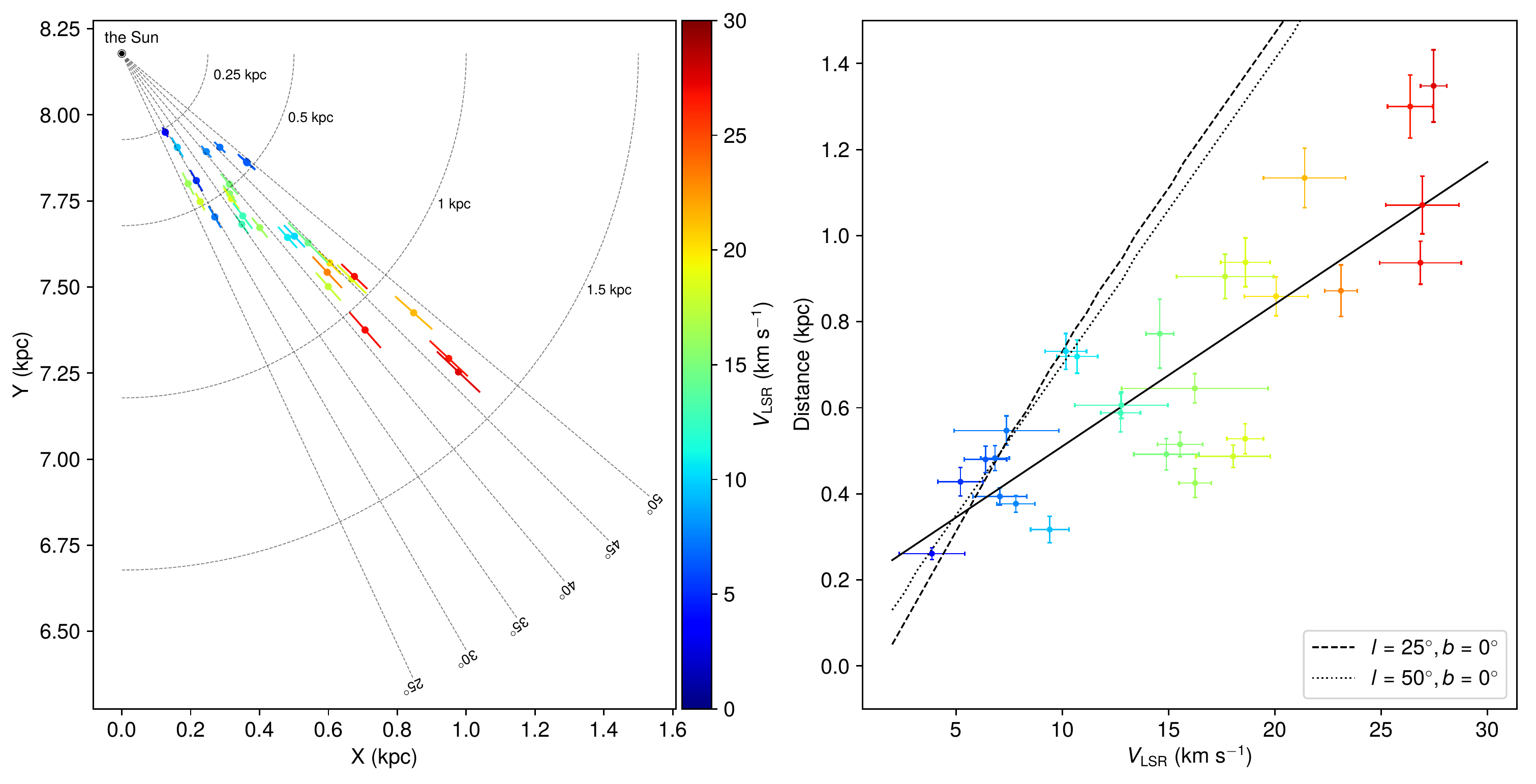}
\caption{Face-on view of 28 molecular clouds with well-determined distances. Colors in both panels represent the magnitude of radial velocities, and the 5\% systematic error is included in the distance error bars. The error of $V_{\rm LSR}$ is the second moment of velocities. With Bayesian analysis and MCMC sampling, we derived a relation of $\frac{D}{\mathrm{[kpc]}}=\left(0.033\frac{V_{\rm LSR}}{[\rm km \ s^{-1}]}+0.180\right)$ between the distance and $V_{\rm LSR}$, and the distance scatter around this linear relationship is about 0.161 kpc. In the right-hand side plot, the dashed and dotted black lines represent the kinematic distance curve \citep{2014ApJ...783..130R} with respect to $V_{\rm LSR}$ toward $l=25^{\circ}$ and $l=50^{\circ}$ in the mid-plane ($b=0^{\circ}$), respectively. A methanol maser source, G035.19-00.74 \citep{2009ApJ...693..419Z} (2.19 kpc at 30 \kms), does not follow this linear relationship but is more close to the kinematic distance. The distance to the Galactic center is 8.178 kpc \citep{2019A&A...625L..10G}. \label{fig:faceon} }
\end{figure*}

Figure \ref{fig:faceon} describes the face-on distribution of the 28 molecular clouds and the relationship between the distance and the radial velocity. The 28 molecular clouds are certainly incomplete, and for each line of sight, only distances to the nearest molecular clouds were derived. However, those clouds show a quite clear linear correlation between the distance and the radial velocity, which is $\frac{D}{\mathrm{[kpc]}}=\left(0.033\frac{V_{\rm LSR}}{\mathrm{[km \ s^{-1}]}}+0.180\right)$. The parameters were derived from Bayesian analysis and MCMC sampling with both errors in radial velocities and distances considered, including the 5\% systematic error in distances.

\begin{figure*}[ht!]
 
\gridline{\fig{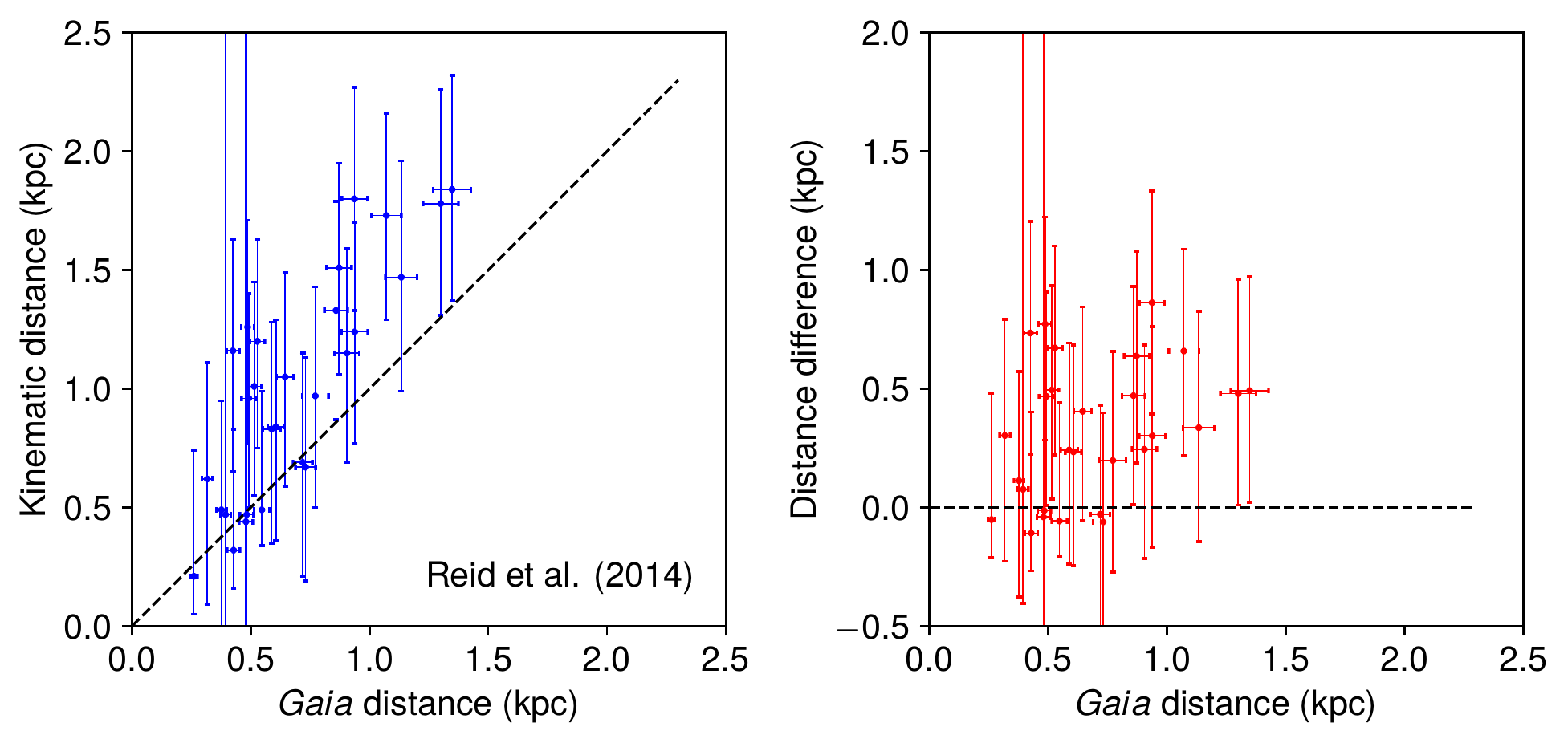}{0.45\textwidth}{(a)} \fig{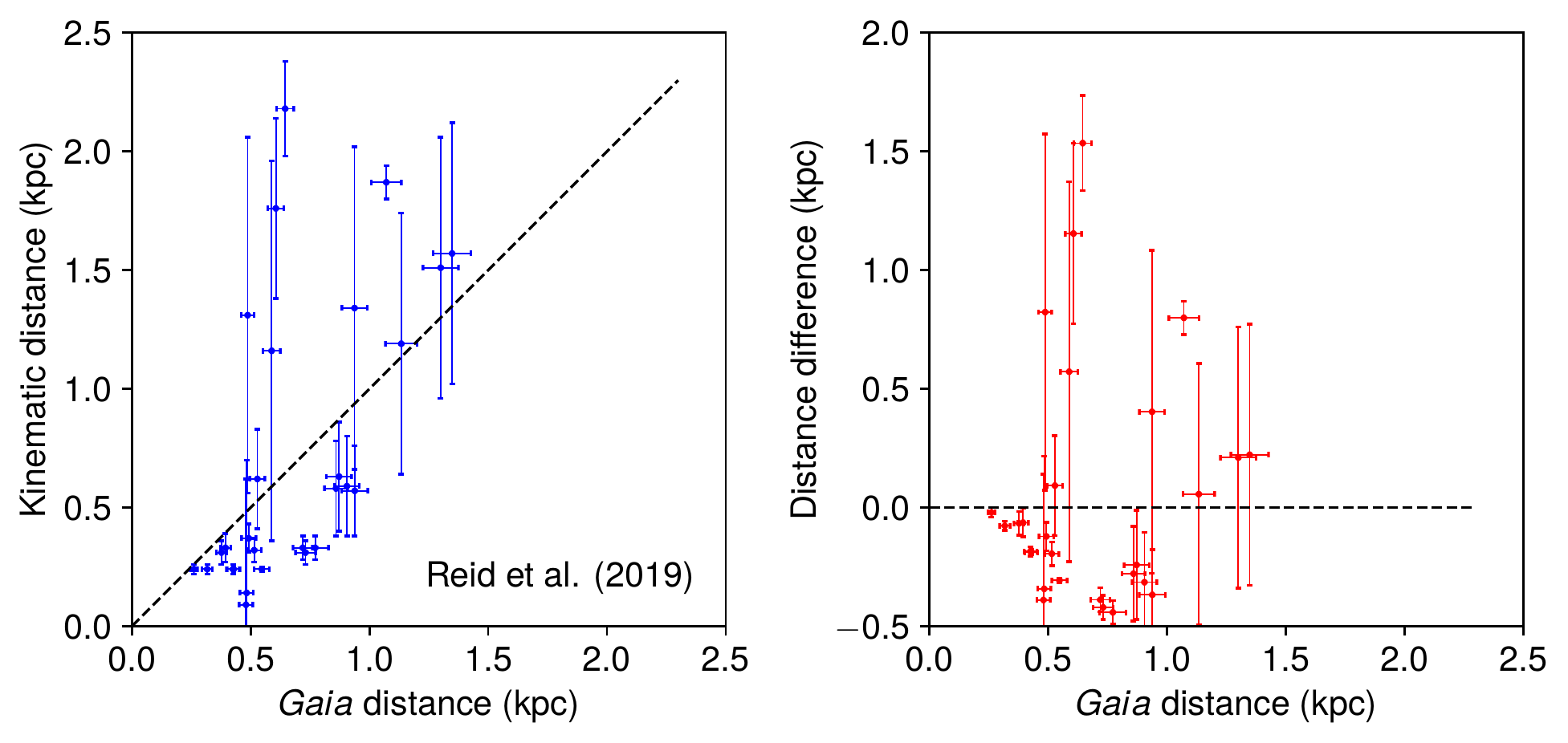}{0.45\textwidth}{(b)}  }
\caption{A comparison of \textit{Gaia} DR2 and near kinematic distances to 28 local molecular clouds in the first Galactic quadrant. Panel (a) shows the comparison results with A5 model of \citet{2014ApJ...783..130R}, while panel (b) shows that with an improved version from \citet{2019ApJ...885..131R}. For molecular clouds at $V_{\rm LSR}\geq 10 $ \kms, the systematic shift of kinematic distances with respect to \textit{Gaia} DR2 distances is 0.43 kpc for panel (a), and 0.15 kpc for panel (b). The 5\% systematic error is included in \textit{Gaia} DR2 distance error bars. \label{fig:comparedis} }
\end{figure*}

%
A comparison of \textit{Gaia} DR2 distances and near kinematic distances is shown in Figure \ref{fig:comparedis}. Kinematic distances have two problems: (1) the errors are too large for molecular clouds with small magnitude of $V_{\rm LSR}$ ($|V_{\rm LSR}|<$ 10 \kms); (2) hard to distinguish the near and far distance ambiguity. We only use near distances because \textit{Gaia} DR2 distances suggest that they are local. For molecular cloud samples with $V_{\rm LSR}\geq10$ \kms, kinematic distances are systematically larger than \textit{Gaia} DR2 distances by about 0.43 kpc \citep{2014ApJ...783..130R}, 0.15 kpc with the updated model of \citet{2019ApJ...885..131R}. This systematic shift indicates that the motion of local molecular clouds ($>1$ kpc) may not precisely follow the Galactic rotational curve. However, most kinematic distances are compatible with \textit{Gaia} DR2 distances within errors. 
 
 
\begin{figure*}[ht!]
 \plotone{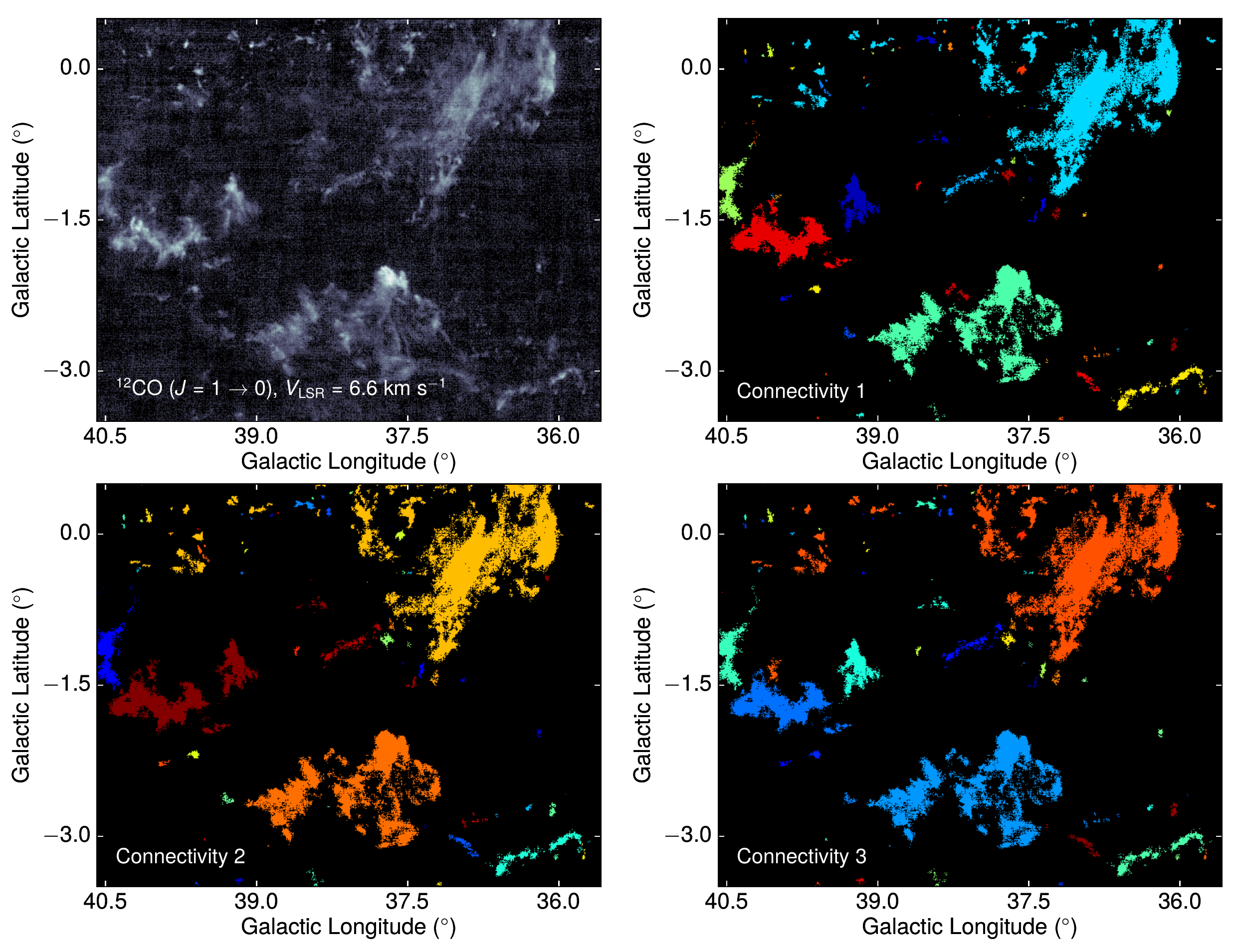}
\caption{Cloud identifications at $V_{\rm LSR} = 6.6$ \kms\ with three connectivity types. The MinPts are 4, 8, and 11 for connectivity 1, 2, and 3, respectively. Colors represent distinct cloud regions, and the CO cutoff is 2$\sigma$ (1.0 K). \label{fig:cloud} }
\end{figure*}


 \begin{figure*}[ht!]
 \plotone{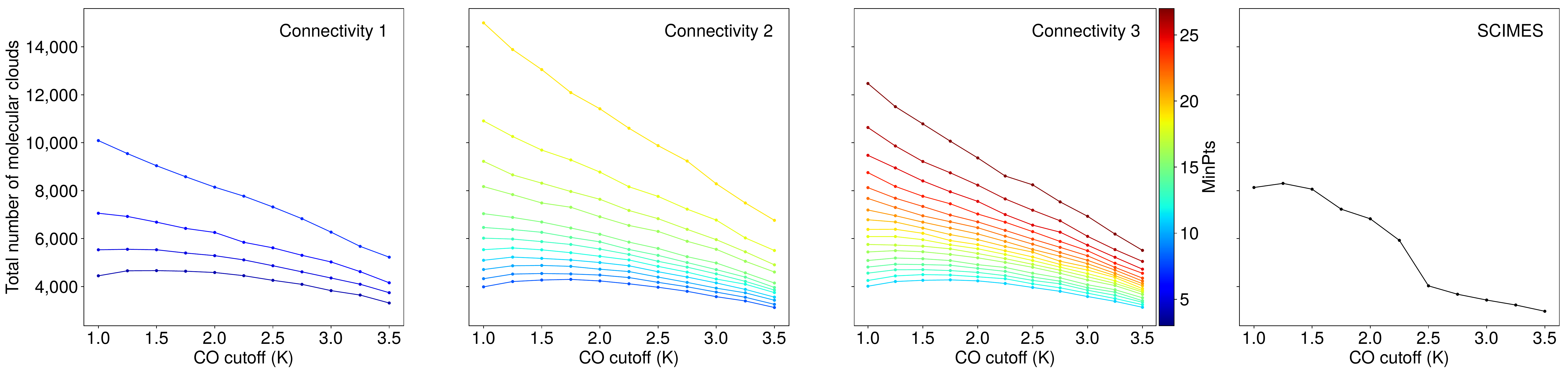}
\caption{Variations of the total number of local molecular clouds with DBSCAN and SCIMES. The cutoff ranges from 1 K (2$\sigma$) to 3.5 K (7$\sigma$), and the color code represents MinPts values. All catalogs are publicly available on the Harvard Dataverse (\href{https://doi.org/10.7910/DVN/6MX6BG}{https://doi.org/10.7910/DVN/6MX6BG}). \label{fig:number} }
\end{figure*}

\subsection{Cloud samples}
\label{sec:cloudN}

Figure \ref{fig:cloud} displays cloud identification cases with DBSCAN at the 1 K cutoff with the smallest MinPts for each connectivity type. The results of the three connectivity types are almost identical, except for a few small-sized molecular clouds, indicating that the cloud identification is robust against connectivity types. It is worth noting that DBSCAN is able to pick up weak emissions with small MinPts values.   

Figure \ref{fig:number} demonstrates the total number of molecular clouds with respect to 11 CO cutoff cases and three connectivity types. For each connectivity type, molecular cloud numbers increase with MinPts. This is because higher MinPts values require strong connections and structures seen with low MinPts would be decomposed by high MinPts. 

Interestingly, with low values of MinPts, molecular cloud numbers decrease toward both high and low CO cutoffs. The decrease of trunk numbers toward high CO cutoffs is reasonable, because molecular clouds with low brightness temperature are washed out and the break of large molecular clouds cannot fully compensate this effect. However, the molecular cloud trunk number also decreases at lower cutoff ends ($<$3$\sigma$). This may be due to three reasons.  First, molecular clouds are incomplete because weak CO emission is overwhelmed by noise. Secondly, the boundary of many molecular clouds is not resolved by the beam size ($\sim$50\arcsec) and small CO cutoff values combine many molecular clouds into single trunks in low brightness temperature regions. Thirdly, it is possible that there are widely distributed diffuse molecular clouds between more condensed molecular clouds, which are all connected under high sensitive observations. 

For high CO cutoffs, SCIMES cluster numbers are close to that produced with DBSCAN (at low MinPts), but increase significantly toward the low CO cutoff end. This is because large molecular clouds were already decomposed at high CO cutoffs, the SCIMES results resemble DBSCAN clusters. However, at low CO cutoffs, and SCIMES splits large dendrogram trunks, producing many more clusters than DBSCAN with low values of MinPts.

\begin{figure*}[ht!]
 \plotone{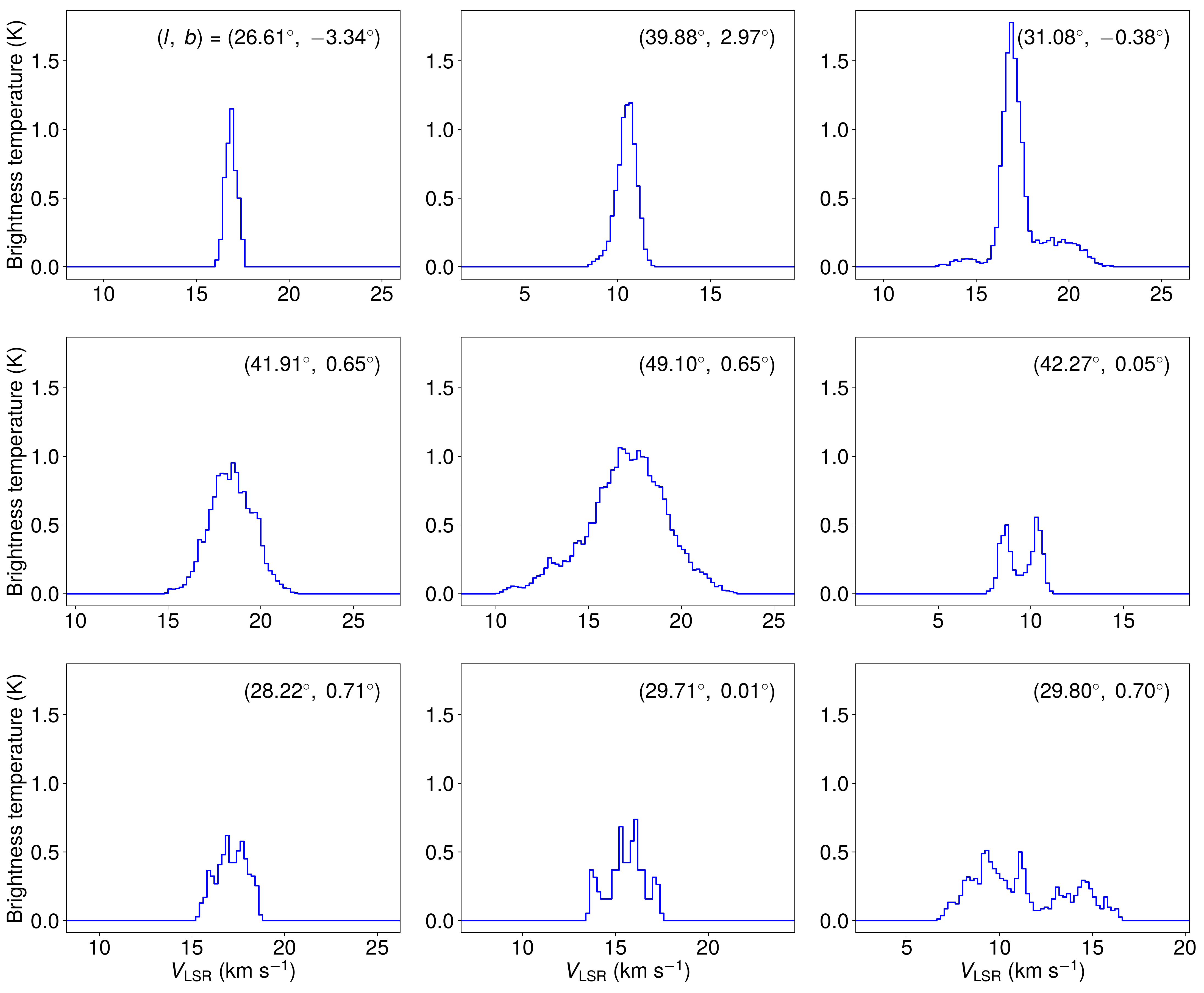}
\caption{Example average spectra of molecular clouds. The molecular clouds are identified with DBSCAN (1 K cutoff, connectivity 1, and MinPts 4), and voxels beyond molecular cloud regions are masked to be 0. \label{fig:spectra} }
\end{figure*}

\begin{figure*}[ht!]

 \gridline{\fig{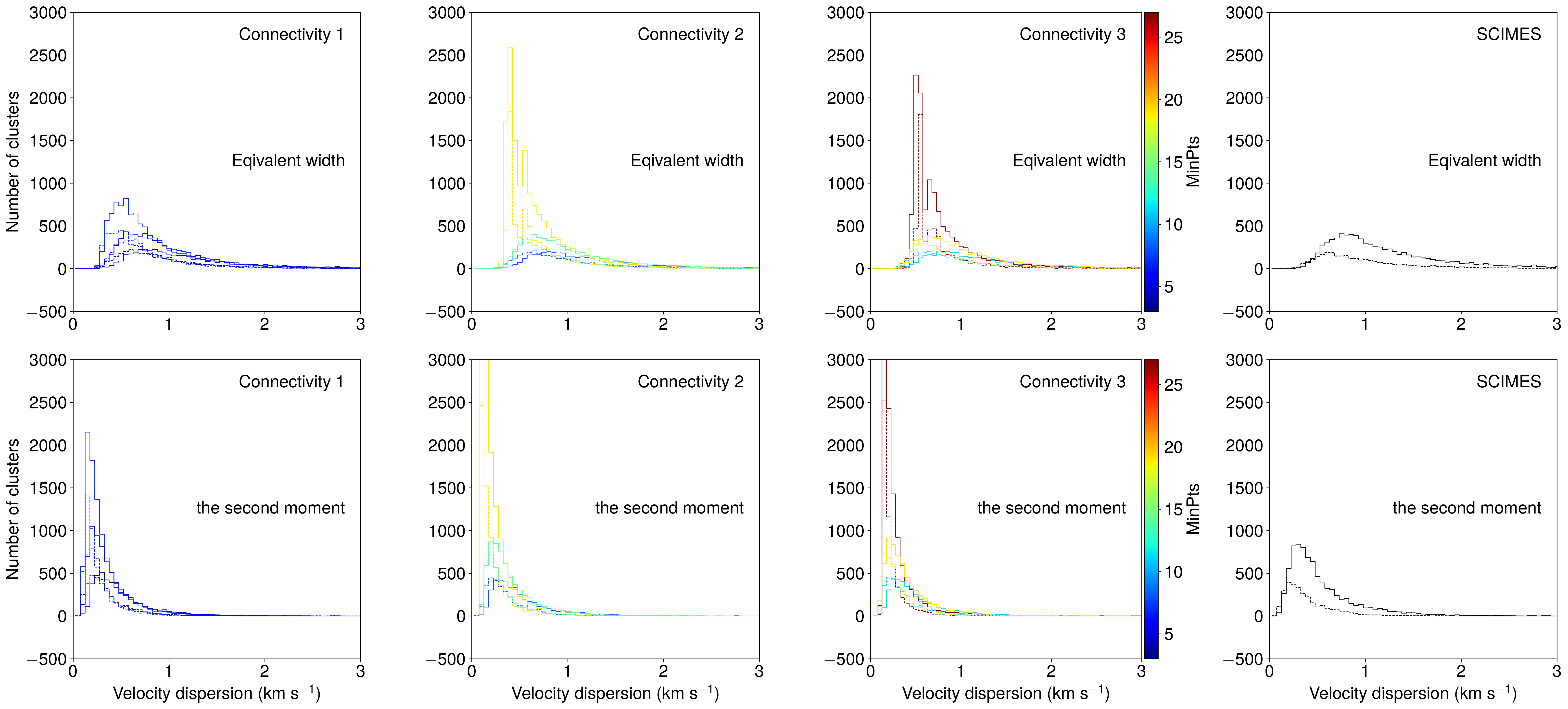}{0.95\textwidth}{(a)}   }
 \gridline{ \fig{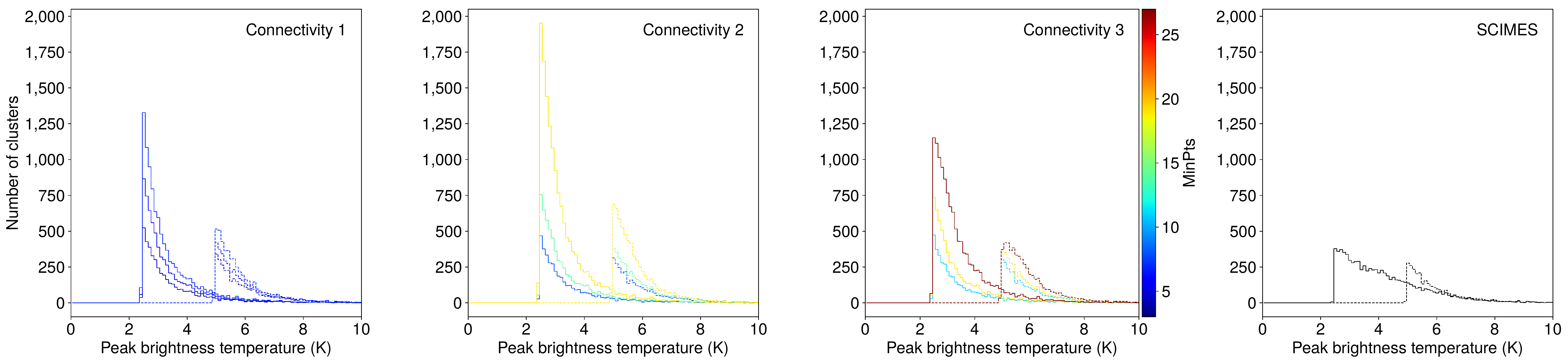}{0.95\textwidth}{(b)}   }

\caption{Distribution of (a) the equivalent linewidth and (b) the peak brightness temperature of molecular clouds. In panel (a), the equivalent linewidth is calculated with the averaged \cofs\ spectrum (see Equation \ref{equ:linewidth}). The solid and dashed lines represent the 1 K (2$\sigma$) and 3.5 K cutoffs, respectively, and to avoid crowding, only three MinPts cases (the smallest, the middle, and the largest ones) are displayed for each cutoff.  \label{fig:velandpeak} }
\end{figure*}

\subsection{Equivalent linewidth and peak distribution}

We now display the distribution of the equivalent linewidth and the peak brightness temperature of molecular clouds. As shown in the averaged spectra of nine representative molecular clouds in Figure \ref{fig:spectra}, many spectra cannot be described with single Gaussian profiles, so we used the equivalent linewidth as a measure of the velocity dispersion. The equivalent linewidth is defined as 
 
\begin{equation}
W_{v}=\frac{1}{ T_{\rm peak}}  \sum_i T_i\Delta v
\label{equ:linewidth}
\end{equation}
where $T_i$ is the brightness temperature of the average spectra in each velocity channel, $\Delta v=0.2$ \kms, and $T_{\rm peak}$ is the peak brightness temperature of the averaged spectra.

The top panel of Figure \ref{fig:velandpeak} describes the equivalent linewidth distribution, and another measurement of the velocity dispersion, the second moment, is also shown as a comparison. The range of the equivalent linewidth is about 0.2-10.0 \kms. For low MinPts cases, the equivalent linewidth shows a similar distribution with that of Gaussian component samples identified by \citet{2020A&A...633A..14R} from the entire Galactic Ring Survey (GRS) \cos\ data using the GaussPy+ algorithm (see Figure 8 therein).  However, with high MinPts, DBSCAN identifies many bright regions as independent molecular clouds, and the number of molecular clouds with small equivalent linewiths increases sharply.

\begin{figure*}[ht!]
  \gridline{\fig{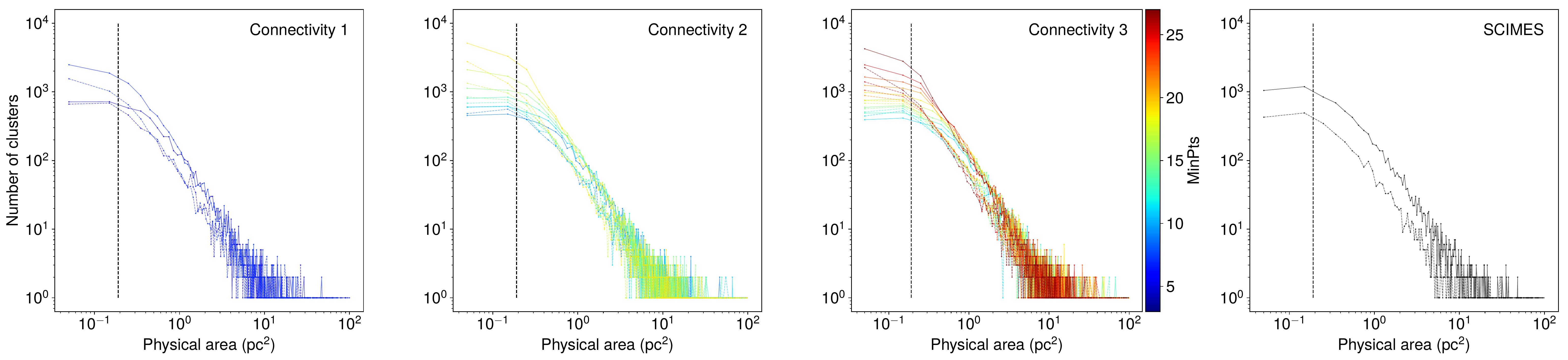}{0.95\textwidth}{(a)} } 
 \gridline{\fig{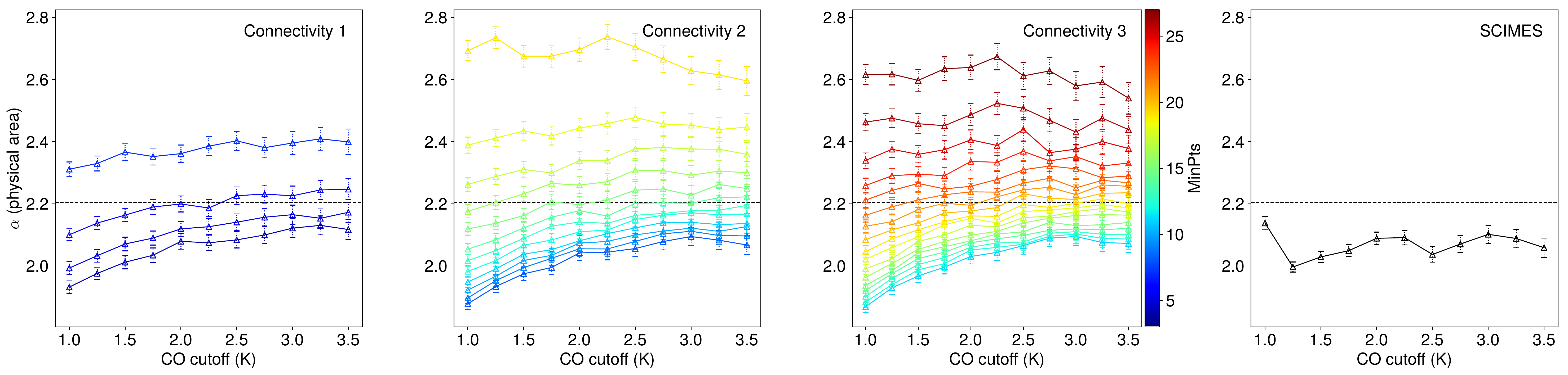}{0.95\textwidth}{(b)} }
\caption{(a) Physical area distribution and (b) power-law index $\alpha$ fitting. To avoid crowding, only odd values of MinPts and two cutoffs (2$\sigma$ in solid lines and 7$\sigma$ in dashed lines) are shown in panel (a), while all cases are displayed in panel (b). Molecular cloud distances are estimated with the distance and radial velocity relationship (see Figure \ref{fig:faceon}). In panel (a), black vertical dashed lines mark the complete threshold of the physical area, 0.19 pc$^2$ (4 pixels at 1.5 kpc). In panel (b), the dashed horizontal black lines represent an $\alpha$ of $2.20\pm0.18$ averaged over all DBSCAN cases. \label{fig:physicalarea} }
\end{figure*}

The peak position of the equivalent linewidth distribution shifts left slightly from 2$\sigma$ (solid lines) to 7$\sigma$ (dashed lines) cutoffs. Two processes are responsible for this shift: (1) the remove of low CO emission in the envelope would decrease the equivalent linewidth and (2) high CO cutoffs break large molecular clouds into ones with small equivalent linewidths, compensating the effect of the first process. Due to the second process, the peak shifting with high CO cutoffs is not evident, and low CO cutoffs produce more molecular clouds with large equivalent linewidths.

The bottom panel of Figure \ref{fig:velandpeak} demonstrates the peak intensity distribution. The peak brightness temperature distribution resembles an exponential distribution above the threshold. As seen in panel (b) of Figure \ref{fig:velandpeak}, observations are incomplete due to the truncation caused by the post selection criteria, and the behavior of the peak brightness temperature toward the lower value end is unknown. CO cutoffs affect the peak distribution significantly, because molecular clouds with low peak intensities are removed by high CO cutoffs and the post selection criteria. The break of large molecular clouds increases the number of molecular clouds with moderate peak brightness temperatures.

\subsection{Physical  area distribution}
\label{sec:area}


In this section, we describe the physical area distribution of molecular clouds. The molecular cloud distances are estimated with the distance and radial velocity relationship (see Figure \ref{fig:faceon}), and the residual distance dispersion (161 pc) is used as the distance error, which is the only error source considered. Typically, the relative error of the physical area is about 50\%. Molecular clouds nearer than 200 pc or farther than 1500 pc are excluded in statistics, due to the uncertainty of extrapolation.

 \begin{figure*}[ht!]
 \plotone{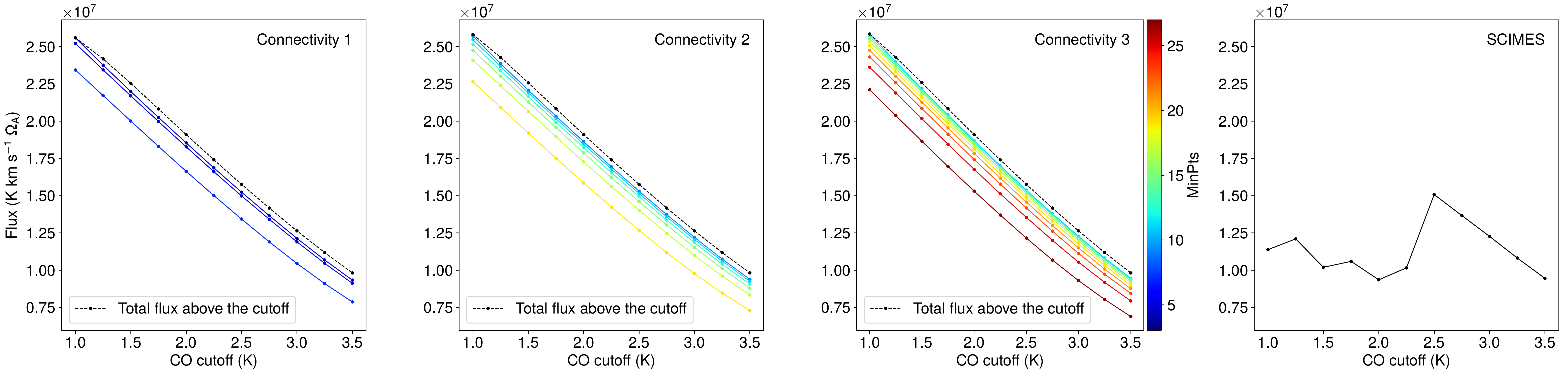}
\caption{Flux reconstructed from DBSCAN and SCIMES. For three connectivity types, black dashed lines represent the total flux, which is the direct sum (multiplied by a factor of 0.2 \kms\ $\Omega_{\rm A}$) of all voxel intensities above a specific cutoff, based on the noise-masked data at the 1 K (2$\sigma$) cutoff and the smallest MinPts for each connectivity type. To avoid crowding, we only plot odd MinPts values. The linear relationship between the total flux and the cutoff is similar among the three connectivity types, and can be approximated by ($-6.5\times10^6$$\frac{\rm cutoff}{[\rm K]}$ + $3.2\times10^7$) $\rm K\ km\ s^{-1}\ \Omega_{A}$, where $\Omega_{\rm A}$ is the solid angle of a single voxel. \label{fig:allflux}  } 
\end{figure*}

The minimum physical area of molecular clouds is about 0.01 pc$^2$, and the maximum physical area is about 1$\times$10$^4$ pc$^2$. The dynamic range of the physical area is $\sim$10$^6$, which enables us to gauge the area distribution robustly. As to the angular area, the smallest molecular cloud only has one beam size (4 pixels), while the largest one occupies more than half of the entire surveyed region. The largest molecular cloud is mainly the Aquila Rift and is contaminated by some molecular clouds from the Perseus arm.

As shown in Figure \ref{fig:physicalarea}, the physical area shows a power-law distribution.  This power-law distribution is insensitive to algorithms and parameters. Even for SCIMES clusters, which do not assign a large portion of the CO emission in the envelope, the area distribution is still present. We fitted the index of each case with a truncated power-law model (with a minimum threshold), and derived the index with Bayesian analysis and MCMC sampling. The complete threshold of the physical area,  $A_{\mathrm{min}}$, is estimated by assuming 4 pixels is the minimum resolved angular area,  and according to the distance range (0.25-1.5 kpc, see \S\ref{sec:distance}), 4 pixels at 1.5 kpc corresponds to 0.19 pc$^2$, i.e.,  $A_{\mathrm{min}}$ = 0.19 pc$^2$. Consequently, only molecular clouds that have physical areas larger than 0.19 pc$^2$ are used in the power-law fitting.


Maser parallax measurements \citep{2019AJ....157..200Z,2019ApJ...885..131R} show that part of the molecular clouds ($-6$ to 30 \kms) are located in the Perseus arm ($>$ 8.8 kpc). The contamination of the Perseus arm molecular clouds may cause systematic errors in the $\alpha$ index. Given the far distances and small filling factors, the angular area of molecular clouds in the Perseus arm is relatively small, and many are locked in the largest molecular cloud, consequently, the contamination of the Perseus arm molecular clouds is negligible.


The power-law distribution is defined as 
\begin{equation}
\mathrm{d}N \propto A^{-\alpha} \mathrm{d} A, 
\end{equation}
where $A$ is the molecular cloud physical area. The normalized probability density function (PDF) for each molecular cloud (no errors considered) is
\begin{equation}
P\left(A_i|\alpha\right) = \left(\alpha-1\right) A_{\mathrm{min}}^{\alpha-1}A_i^{-\alpha}, \label{eq:pdfraw}
\end{equation}
 where $A_i$ is the physical area of a molecular cloud. However, if the error of $A_i$, $\Delta A_i$, were considered, the PDF would be the convolution \citep{2009MNRAS.397..495K} of the error distribution with Equation \ref{eq:pdfraw}, which is  

\begin{equation}
P\left(A_i,\Delta A_i|\alpha\right) = \int \frac{1}{\Delta A_i\sqrt{2\pi}}\exp\left(-\frac{1}{2}\left(\frac{A_i-x}{\Delta A_i}\right)^2\right) P\left(A_i-x|\alpha\right) dx \label{eq:pdferror}, 
\end{equation}
where x is the $A_i$ error, whose PDF is assumed to be Gaussian with a mean and standard deviation of $A_i$ and  $\Delta A_i$, respectively. The total probability is the product of PDFs of all physical areas, and the involvement of the integration, which has no closed-form expression, makes the  MCMC sampling slow.  Consequently, we calculated 50 MCMC sampling chains, each of which contains 40 thinned samples (every 10) and extra 10 burn-in (the first 10 thinned samples were discarded). Consequently, the total sample number is 2000, whose mean and standard deviation correspond to the mean and error of $\alpha$.

 Panel (b) of Figure \ref{fig:physicalarea} displays the $\alpha$ derived with different CO cutoffs and algorithms. The standard deviation of each $\alpha$ derived from MCMC sampling for each case is approximately equal ($\sim$0.03), so we took the unweighted mean of all $\alpha$ as the estimated value (2.20), and convolved the MCMC error to the unweighted standard deviation of all $\alpha$ (only with DBSCAN) as the total error (0.18), i.e., the estimated $\alpha$ is 2.20$\pm$0.18. SCIMES results are not included. The $\alpha$ shows a systematic variation with respect to CO cutoffs and minPts, and roughly, it increases slightly toward high CO cutoffs and large minPts values. Interestingly, both high CO cutoffs and large minPts values correspond to regions with bright \cofs\ emission, i.e., diffuse molecular clouds may have different statistical properties with dense molecular clouds. 
\begin{figure*}[ht!]
  \gridline{\fig{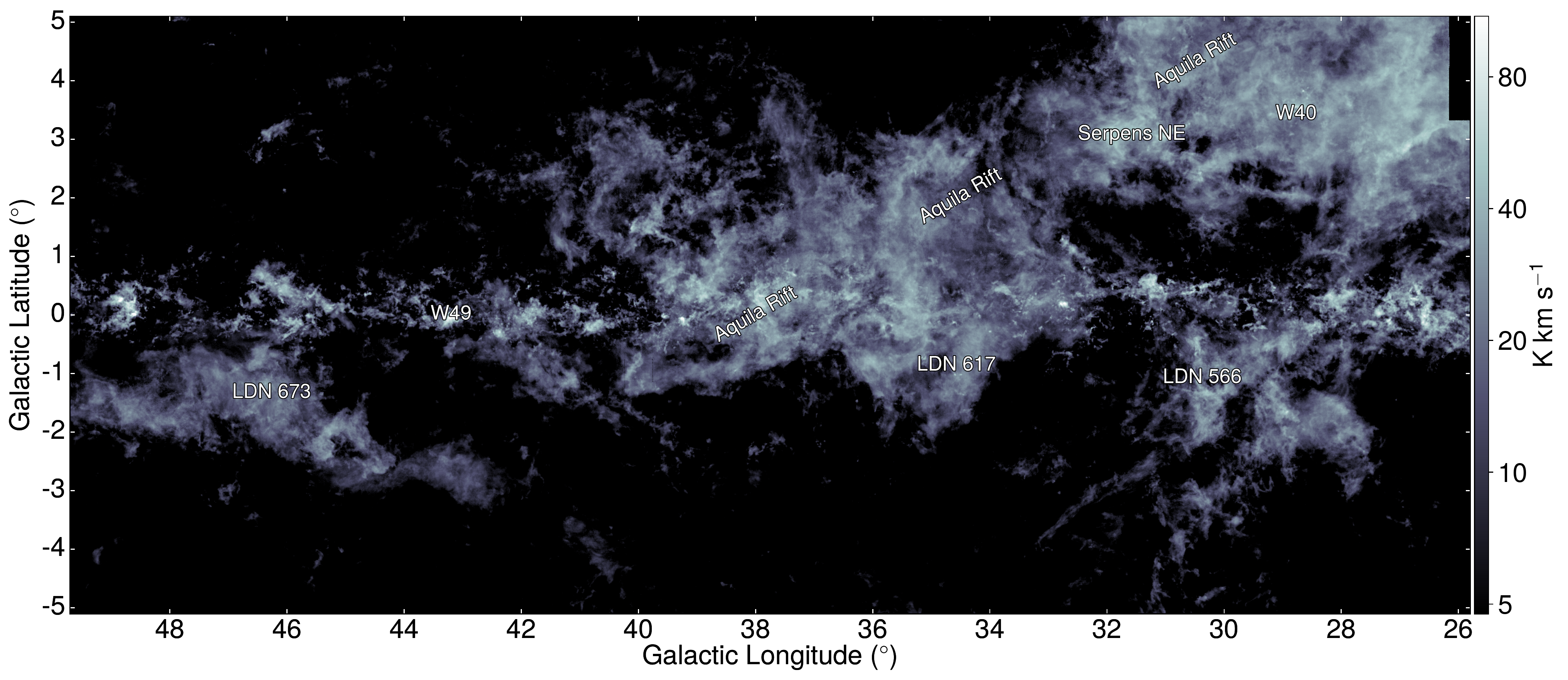}{0.95\textwidth}{(a)} } 
 \gridline{\fig{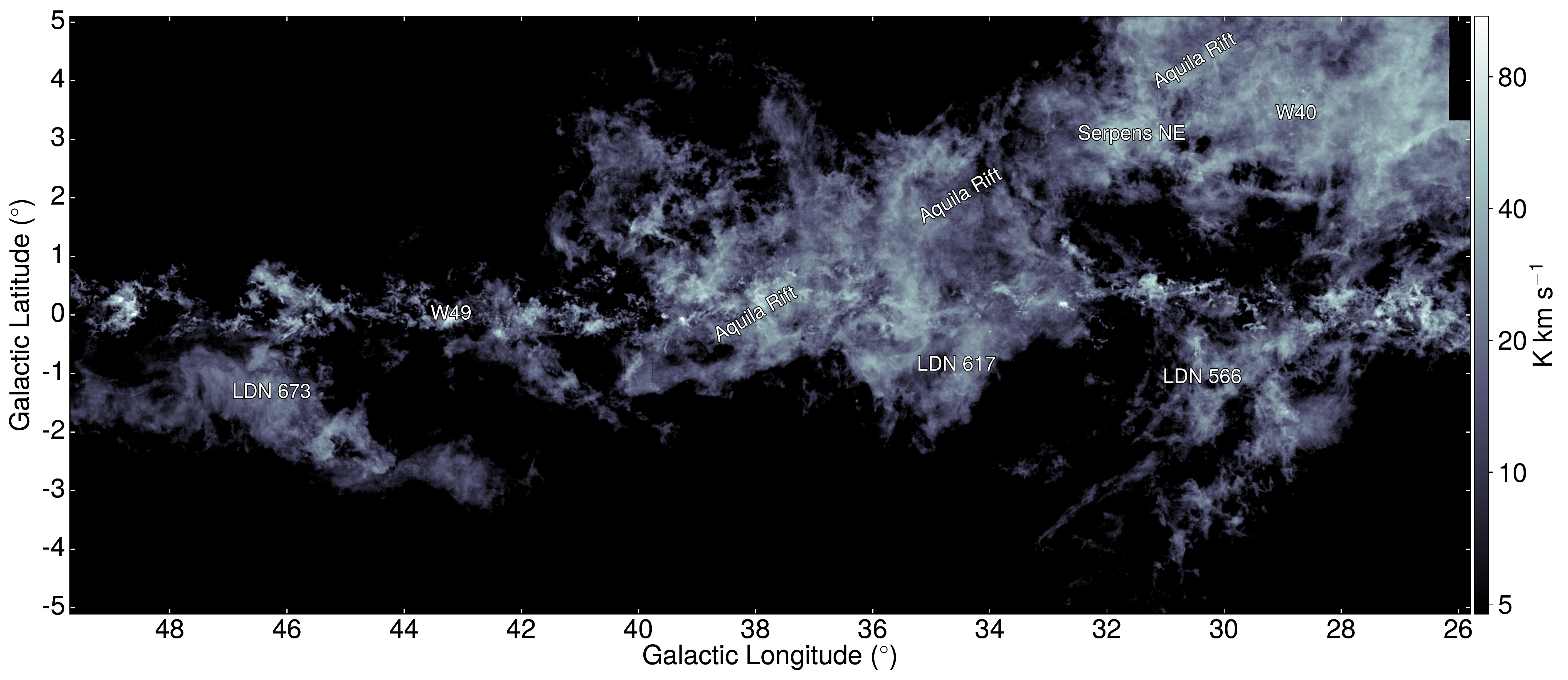}{0.95\textwidth}{(b)} }
\caption{ (a) The integrated intensity map of all molecular clouds identified with DBSCAN (minPts = 4 and connectivity = 1) at 2$\sigma$ level. (b) The integrated intensity map of the largest molecular cloud in panel (a). \label{fig:intcloud} }
\end{figure*}

\subsection{Mass distribution} 

In addition to the physical area, we examined the distribution of the cloud mass. The mass is derived by assuming a $^{12}$CO-to-H$_2$ mass conversion factor of X = $2.0\times 10^{20}$ cm$^{-2}$ $\rm(K\  km\ s^{-1})^{-1}$ \citep{2013ARA&A..51..207B} (see Table \ref{Tab:cloudDis}). The minimum and maximum cloud mass is about 0.05 and $5\times10^5$ \msun, and the relative mass error is typically 50\%.


We first examined the total flux reconstructed from the molecular clouds identified with DBSCAN and SCIMES. As demonstrated in Figure \ref{fig:allflux}, for each cutoff, the total flux of \cofs\ is calculated based on the noise-masked spectral data with the 1 K (2$\sigma$) cutoff and the smallest MinPts values for each connectivity type. For a specific cutoff, the total flux is the sum of all voxel (noise-masked) brightness temperatures multiplied by 0.2 \kms\ $\Omega_{\rm A}$, where $\Omega_{\rm A}$ is the solid angle of a single voxel.   

\begin{figure*}[ht!]
  \gridline{\fig{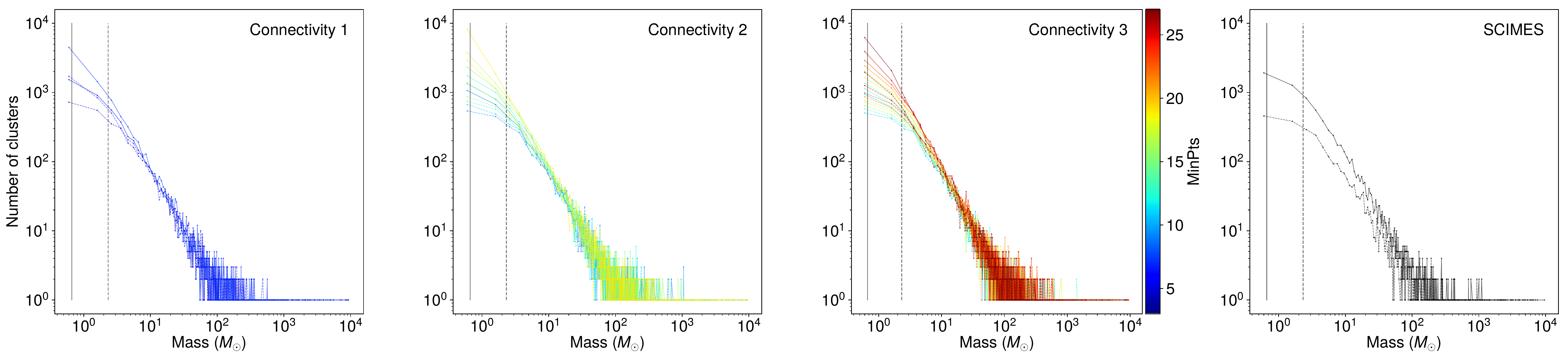}{0.95\textwidth}{(a)} } 
 \gridline{\fig{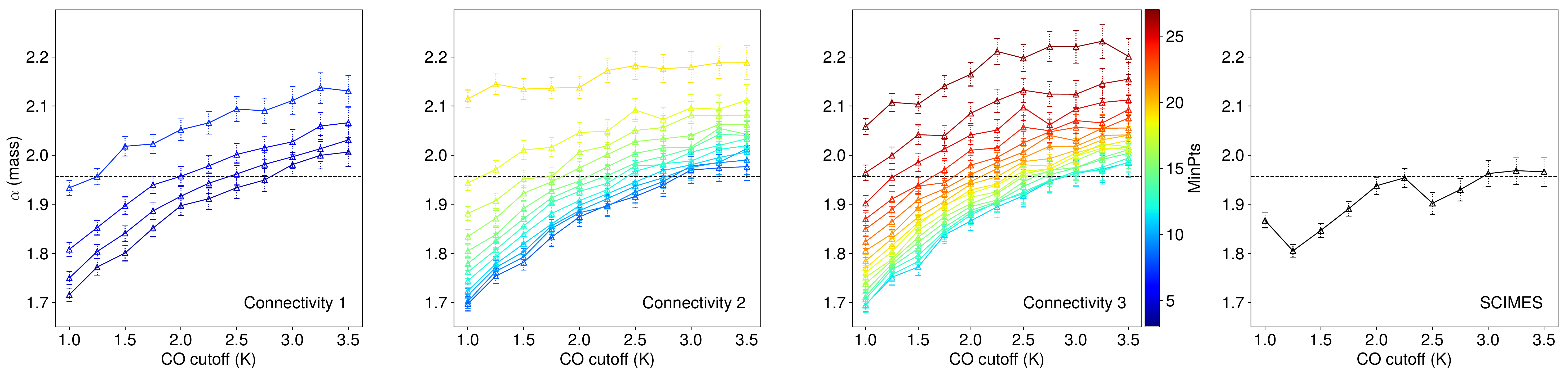}{0.95\textwidth}{(b)} }
 \caption{(a) mass distribution and (b) power-law index $\alpha$ fitting. In panel (a), the solid and dashed vertical lines mark the complete mass threshold which is calculated with the minimum flux (16$\times$cutoff$\times$0.2 \kms) at 1.5 kpc, where 16 is the minimum voxel number required.  \label{fig:mass} }
\end{figure*}

Strikingly, as shown in Figure \ref{fig:allflux}, the total flux of \cofs\ increases approximately linearly with the decrease of the CO cutoff, and there is no sign of completeness at the low CO cutoff end. The total flux can be approximated by ($-6.5\times10^6$$\frac{\rm cutoff}{[\rm K]}$+$3.2\times10^7$) $\rm K\ km\ s^{-1}\ \Omega_{A}$. This indicates that despite being weak ($<$2 K, 4$\sigma$), the CO emission around molecular clouds collectively has the same order of magnitude as strong emission ($\geq$2 K) due to the large volume. Clearly, observations with a sensitivity of 0.5 K are still incomplete for local molecular clouds ($<$ 1.5 kpc).

DBSCAN recovers most of the flux in cases with small MinPts values, but SCIMES misses a significant fraction (about 55\%) at the lower CO cutoff end ($\leq$ 2.5 K). The maximum flux of molecular clouds is about $2.4\times10^7$ K \kms\ $\Omega_{\rm A}$, and the dynamic range of the flux is about 10$^6$  (the minimum flux is about 4 K \kms\ $\Omega_{\rm A}$). DBSCAN shows that the molecular cloud that has the highest flux occupies about 92\% percent of the total flux at the 2$\sigma$ cutoff and decrease to 42\% at the 7$\sigma$ cutoff. Interestingly, the statistical properties of molecular clouds can be revealed by a large number of small independent structures, which collectively occupy a tiny portion of the total flux in PPV space.  

 In Figure \ref{fig:intcloud}, we display the integrated intensity map of all DBSCAN (minPts = 4 and connectivity = 1) molecular clouds at 2$\sigma$ cutoff. As a comparison, we display the integrated intensity map of the largest molecular clouds, which contains about 92\% of the total molecular cloud flux. The main component of the largest molecular cloud is the Aquila Rift, and evidently,  it contains a fraction of emission from the Perseus arm near the Galactic mid-plane.





Panel (a) of Figure \ref{fig:mass} displays the mass distribution of molecular clouds identified with DBSCAN and SCIMES. The \ref{fig:mass}  shows a power-law distribution, and we fitted the power-law index with the same procedure that used on the physical area, which is displayed in panel (b) of Figure \ref{fig:mass}. The complete threshold of the mass is estimated  with the minimum flux (16$\times$cutoff$\times$0.2 \kms) at 1.5 kpc.    The systematic variation of the mass power-law index with respect to CO cutoffs and minPts values is more evident than that of the physical area, and the average power-law index, $1.96\pm0.11$, of the mass distribution is slightly flatter than that of the physical area.



\section{Discussion}
\label{sec:discuss}
In this section, we compare our results with previous studies, including molecular cloud properties and their distances.




\subsection{DBSCAN versus HDBSCAN}
\label{sec:hdbscan}

In addition to DBSCAN, HDBSCAN (Hierarchical Density-Based Spatial Clustering of Applications with Noise), an improvement version of DBSCAN, is also able to perform  clustering. We compared these two algorithms and conclude that HDBSCAN is not suitable to identify consecutive structures in PPV space. Consequently, we used DBSCAN instead of HDBSCAN.  

\begin{figure*}[ht!]
 \plotone{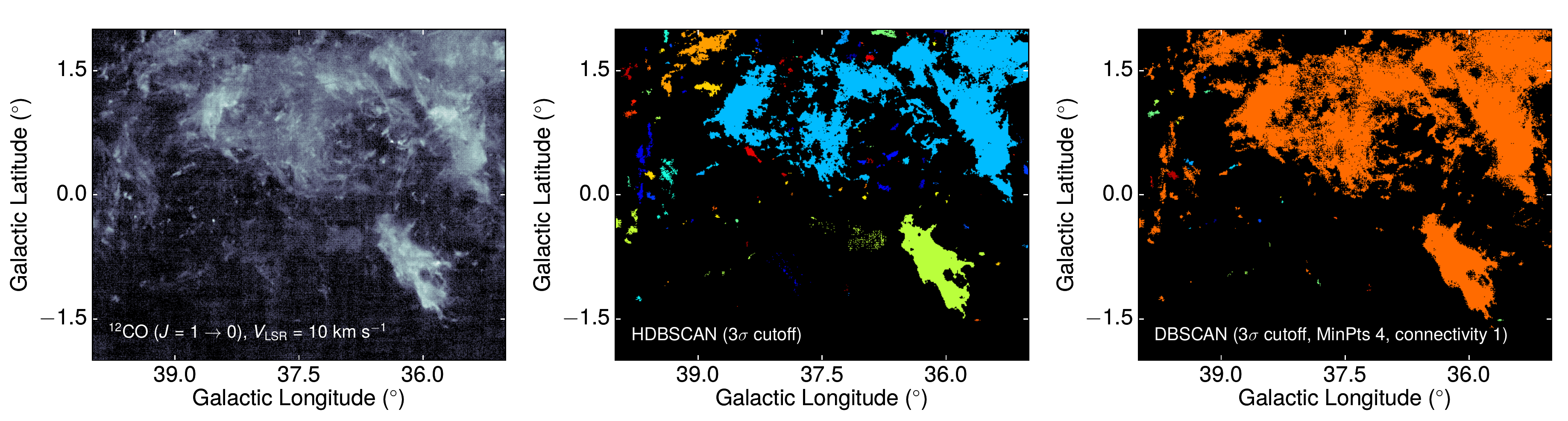}
\caption{A comparison of HDBSCAN and DBSCAN clusters. \label{fig:hdbscan} }
\end{figure*}

HDBSCAN adjusts the value of $\epsilon$ according to the density of points, i.e., the definition of molecular clouds is not uniform and changes with regions. As shown in Figure \ref{fig:hdbscan}, HDBSCAN misses significant flux in dense regions due to high point densities (corresponding to small $\epsilon$), while in sparse regions, HDBSCAN collects loosely bound (corresponding to large $\epsilon$)  points as clusters. In addition, we found that HDBSCAN cluster results depend on the $l$-$b$-$V$ range of the input data cube, which is also a common problem of SCIMES.

\subsection{Molecular cloud distances}
 
In this section, we compare \textit{Gaia} DR2 distances with previous studies.

Eight maser sources \citep{2009ApJ...693..419Z, 2013ApJ...775...79Z, 2019AJ....157..200Z} with distances derived from trigonometric parallaxes are located in the $l$-$b$-$V$ space of local molecular clouds, but they are all too far to be local components. Seven of those masers are located in the Perseus arm ( $0 \leq V_{\rm LSR}  \leq 30 $ \kms\ and $D>$ 8.85 kpc), and the remaining one, G035.19-00.74 \citep{2009ApJ...693..419Z}, belongs to the Sagittarius arm.  Due to the far distance or heavy foreground extinction, none of the eight maser sources has extinction distance measured. The distance of G035.19-00.74 is about 2.19 kpc (at 30 \kms), which shows large deviations from the linear relationship between the local distance and the radial velocity, but the distance of G035.19-00.74 is more close to the kinematic distance.  Consequently, although in the same $l$-$b$-$V$ space, some molecular clouds are not local.


G029.6+03.7 contains a nearby \HII\ region, W40, which is known in the Serpens molecular cloud. The \textit{Gaia} DR2 distance of G029.6+03.7 is $547_{- 39}^{+ 39}$ pc, while the VLBA parallax measurements \citep{2017ApJ...834..143O} show that the distance of W40 is 436$\pm$9 pc. However, the W40 region is complicated, and no close off-cloud regions are available. As shown in panel (a) of Figure \ref{fig:twoclouds}, the selected region is about 5\deg\ to the left of W40. At a distance of 436 pc, 5\deg\ corresponds to about 40 pc, suggesting that the 100-pc difference is reasonable. This also shows that the size of molecular clouds can be as large as 100 pc along the line of sight, and small subregions may not be able to represent averaged distances to large molecular clouds. This distance dispersion is consistent with the results of \citet{2019ApJ...878..111H} and \citet{2020A&A...633A..51Z}.

\subsection{Number of molecular cloud samples}

An important step toward understanding molecular clouds is to obtain a complete census of their population. However, unlike stars, molecular clouds are extended sources, and their total number depends on tracers, definitions, and observation qualities.   

In the surveyed 239 deg$^2$ region,  the number of clouds per square degree is about 19 according to the results of connectivity 1 (MinPts 4). \citet{2019MNRAS.483.4291C} identified about 85000 clouds in the first Galactic quadrant ($10.25^\circ<l<17.5^\circ$ and $|b|\leq0.5^\circ$, $17.5^\circ<l<50.25^\circ$ and $|b|\leq 0.25^\circ$, and   $50.25^\circ<l<55.25^\circ$ and $|b|< 0.5^\circ$) from the JCMT \cotss\ High-Resolution Survey \citep[COHRS, ][]{2013ApJS..209....8D}, and  5229 of those 85000 clouds are in the local  $l$-$b$-$V$ space. \citet{2019MNRAS.483.4291C} found many more clouds (in a much smaller region) because they split large trunks into small components with SCIMES, and furthermore, \cotss\ requires high temperature environments to excite and is less crowded in $l$-$b$-$V$ space.




\citet{2016ApJ...822...52R} provided a uniform catalog of 1064 massive molecular clouds based on the CfA-U.Chile \cof\ survey \citep{2001ApJ...547..792D}, and the technique they used is the dendrogram. Only 16 of those molecular clouds are located in the $l$-$b$-$V$ space of local molecular clouds.  However, using the same data set but only focus on the Galactic plane ($|b|<5^\circ$), \citet{2017ApJ...834...57M}  found 8107 molecular clouds using an alternative algorithm, which combines the hierarchical cluster identification and Gaussian decomposition, and 380 molecular clouds are located in the local  $l$-$b$-$V$ space. The large beam size (about 8\arcmin) of the CfA-U.Chile CO survey overwhelms many small-sized ones.

According to the DBSCAN and SCIMES results, DBSCAN is useful for detecting consecutive structures in PPV space, while SCIMES is capable of splitting large molecular clouds into moderate ones. In the first Galactic quadrant, DBSCAN collects a large portion of PPV voxels into single large molecular clouds with low minPts values, while SCIMES splits this large structure into small-sized ones, producing many more molecular clouds than DBSCAN. However, in the second and third Galactic quadrant, where molecular clouds are not crowded in PPV space, DBSCAN and SCIMES results would be similar.

The molecular cloud samples detected by the MWISP CO survey are still incomplete. The turnover of molecular cloud numbers with small MinPts (see Figure \ref{fig:number}) near 3$\sigma$ may be due to low signal-to-noise ratios. By extrapolating the linear relation between the CO cutoff and total flux described in Figure \ref{fig:allflux} to 0 K, the maximum flux is estimated to be $3.23\times10^7$ $\rm K\ km s^{-1}\ \Omega_{A}$ at the sensitivity of the PMO 13.7-m telescope, meaning that the 2$\sigma$ cutoff collects about 80\% of the total flux (70\% for the 3$\sigma$ cutoff). Compared to higher sensitive telescopes, the PMO 13.7-m telescope may still miss a significant part of the flux.


\subsection{The physical area and mass distribution}
\label{sec:areadiscuss}

The spectra of the physical size \citep{2019MNRAS.483.4291C} and the mass \citep{2016ApJ...822...52R} of molecular clouds are charactered with power-law distributions. However, as demonstrated by DBSCAN results, the physical area and mass distributions are generally conform with the power law, but their indices vary systematically with the DBSCAN parameters and the CO cutoffs.

 The statistics of DBSCAN molecular clouds show significant difference with previous results. Averagely, the power-law index of the physical area is about -2.20, and the corresponding index of the molecular cloud size is about -3.40, which is steeper than the value -2.8 obtained by \citet{2019MNRAS.483.4291C} with SCIMES. The power-law index of the mass spectrum is about -1.96, which is moderate compared with that (-2.2) found by \citet{2016ApJ...822...52R} with a uniform catalog built on a large-scale Galactic \cof\ survey \citep{2001ApJ...547..792D} using the dendrogram and with that (-1.70) found by \citet{2019MNRAS.483.4291C} with \cotss\ line using SCIMES. 

The variation of the power-law index can be attributed to multiple causes. Apart from the tracing transition lines, algorithm parameters, the distance errors, and regions, another factor that may significantly affect the molecular cloud size and mass distribution is the filling factor. The filling factor is related to the sensitivity and resolution of the telescope and some intrinsic properties of molecular clouds, such as the column density distribution and the peak brightness temperature. We would examine the  distance and the filling factor effect once we have collected sufficient molecular cloud samples with accurate distances derived from the \textit{Gaia} stellar parallax and extinction measurements.

\section{Summary}
\label{sec:summary}

We used the DBSCAN algorithm to decompose the \cofs\ spectral cube of local components ($-$6 to 30 \kms) in the first Galactic quadrant into molecular cloud individuals, and investigated the statistical properties and distances of molecular clouds. We define molecular clouds as independent consecutive structures in $l$-$b$-$V$ space, which is robust against the criteria of DBSCAN. At the 1 K (2$\sigma$) emission level and with small MinPts, the number of local molecular clouds per square degree is about 19. For CO cutoffs less than 2 K, SCIMES discard about 55\% of the total flux in order to split large dendrogram trunks that are connected by diffuse CO emission. 


We derived distances to 28 molecular clouds, most of which have their distances accurately  determined for the first time. Distances to molecular clouds are in the range of 250-1500 pc, and  the distance shows  a linear relationship  with the radial velocity. \textit{Gaia} DR2 distances indicate that the kinematic distances may be systematically larger for local molecular clouds in the mapped region.

The linear relationship between the cutoff and the total flux indicates a completeness of about 80\% for the flux collected from local molecular clouds. The largest molecular cloud has an area of $\sim$130 deg$^2$, occupying 92\% of the total flux. The physical area of molecular clouds shows a power-law distribution with an index of about $-2.20\pm$0.18, which changes slightly with cutoffs. The molecular cloud mass shows a power-law distribution as well, and the index is about $-1.96\pm$0.11.

\acknowledgments

We thank D. Colombo for his thorough explanations for SCIMES and thank Sam McSweeney for his careful proofreading. We would also like to show our gratitude to other members of the MWISP group, Ye Xu, Hongchi Wang, Zhibo Jiang, Xuepeng Chen, Yiping Ao, Xin Zhou, Min Wang, Shaobo Zhang,  and Zhiwei Chen, for their useful discussions. We are also immensely grateful to observer assistants---Kun Yan, Min Wang, Dengrong Lu, and Jixian Sun---at PMO Qinghai station for their long-term observation supports. This work was sponsored by the Ministry of Science and Technology (MOST) Grant No. 2017YFA0402701, Key Research Program of Frontier Sciences (CAS) Grant No. QYZDJ-SSW-SLH047, National Natural Science Foundation of China Grant No. 11773077,  and China Postdoctoral Science Foundation under Grant No. 2018M642354. 

%

\vspace{5mm}
\facilities{PMO 13.7-m, \textit{Gaia}. }


\software{astropy \citep{2013A&A...558A..33A},   
          Miriad \citep{1995ASPC...77..433S}, 
          astrodendro,
          scipy
          }

 \bibliographystyle{aasjournal}
 \bibliography{refGAIADIS}





%

\end{document}